\documentclass[journal=jpclcd,manuscript=letter,layout=twocolumn]{achemso}

\usepackage[version=3,version=4]{mhchem} 
\usepackage{graphicx} 
\usepackage{adjustbox}
\usepackage[dvipsnames]{xcolor}
\usepackage{siunitx}
\usepackage[fontsize=10pt]{scrextend}
\usepackage{amssymb} 
\usepackage{caption}
\usepackage{subcaption}
\usepackage{xfrac} 
\usepackage{ulem}
\DeclareGraphicsExtensions{.bmp,.jpg,.pdf,.tif}
\SectionNumbersOff

\makeatletter
\newcommand{\jpcl@sect}[1]{\refstepcounter{section}\par\medskip\noindent\ignorespaces}
\newcommand{\jpcl@sectstar}[1]{\par\medskip\noindent\ignorespaces}
\renewcommand\section{\@ifstar\jpcl@sectstar\jpcl@sect}

\newcommand{\jpcl@subsect}[1]{\refstepcounter{subsection}\par\medskip\noindent\ignorespaces}
\newcommand{\jpcl@subsectstar}[1]{\par\medskip\noindent\ignorespaces}
\renewcommand\subsection{\@ifstar\jpcl@subsectstar\jpcl@subsect}

\newcommand{\jpcl@subsubsect}[1]{\refstepcounter{subsubsection}\par\smallskip\noindent\ignorespaces}
\newcommand{\jpcl@subsubsectstar}[1]{\par\smallskip\noindent\ignorespaces}
\renewcommand\subsubsection{\@ifstar\jpcl@subsubsectstar\jpcl@subsubsect}
\makeatother

\usepackage{comment}

\usepackage{multirow} 

\usepackage{booktabs}   
\usepackage{siunitx}    
\usepackage{tabularx}   
\usepackage{adjustbox}  
\definecolor{blue2}{RGB}{21,114,161}
\definecolor{turquesa1}{RGB}{178, 64, 128}

\newcommand*\MApbi{MAPbI$_3$}
\newcommand*\MApbbr{MAPbBr$_3$}
\newcommand*\FApbi{FAPbI$_3$}
\newcommand*\ma{CH$_3$NH$_3^+$}
\newcommand*\fa{HC(NH$_2$)$^{+}_2$}

\newcommand{\hb}[2]{#1$-$H$\cdots$#2}

\newcommand{\N}{\mathrm{N}}

\newcommand{\Y}{\mathrm{Y}}

\newcommand*\fapifapb{\ce{(FAPbI3)_{7/8}(FAPbBr3)_{1/8}}}
\newcommand*\fapimapi{\ce{(FAPbI3)_{7/8}(MAPbI3)_{1/8}}}
\newcommand*\fapimapb{\ce{(FAPbI3)_{7/8}(MAPbBr3)_{1/8}}}
\newcommand*\fapicspb{\ce{(FAPbI3)_{7/8}(CsPbBr3)_{1/8}}}
\newcommand*\fapi{\ce{FAPbI3}}
\newcommand*\mapi{\ce{MAPbI3}}
\newcommand*\fapb{\ce{FAPbBr3}}
\newcommand*\mapb{\ce{MAPbBr3}}

\author{Liz Camayo-Gutierrez}
\affiliation[utem]
{Departamento de F\'isica, Facultad de Ciencias Naturales, Matem\'atica y del Medio Ambiente (FCNMM), Universidad Tecnol\'ogica Metropolitana, Jos\'e Pedro Alessandri 1242, \~Nu\~noa 7800002, Santiago, Chile}
\author{Javiera Ubeda}
\affiliation[utem]
{Departamento de F\'isica, Facultad de Ciencias Naturales, Matem\'atica y del Medio Ambiente (FCNMM), Universidad Tecnol\'ogica Metropolitana, Jos\'e Pedro Alessandri 1242, \~Nu\~noa 7800002, Santiago, Chile}
\author{Ana L. Montero-Alejo}
\email{amonteroa@utem.cl}
\affiliation[utem]
{Departamento de F\'isica, Facultad de Ciencias Naturales, Matem\'atica y del Medio Ambiente (FCNMM), Universidad Tecnol\'ogica Metropolitana, Jos\'e Pedro Alessandri 1242, \~Nu\~noa 7800002, Santiago, Chile}
\author{Ricardo Grau-Crespo}
\affiliation{School of Engineering and Materials Science, Queen Mary University of London, Mile End Road, London E1 4NS, UK}
\author{Eduardo Men\'endez-Proupin}
\email{emenendez@us.es}
\affiliation{Departamento de Física Aplicada I, Escuela Politécnica Superior, Universidad de Sevilla, Seville E-41011, Spain}

\title{Thermodynamic Stability and Hydrogen Bonds in Mixed Halide Perovskites}

\abbreviations{MAPI, ...}

\begin{document}


\begin{tocentry}
\includegraphics[width= 5 cm,keepaspectratio]{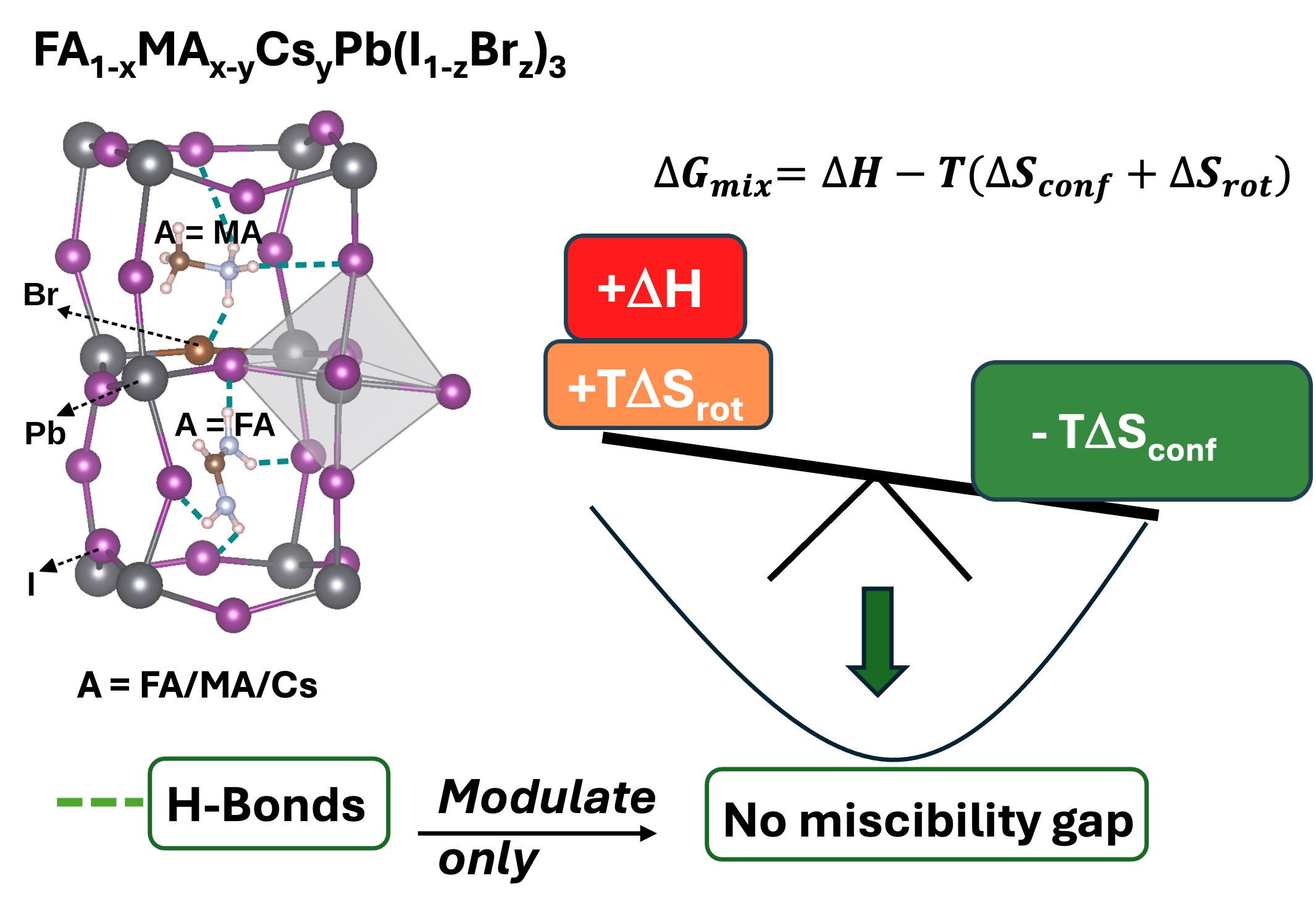}
\end{tocentry}

\begin{abstract}
The stability of mixed halide perovskites against phase separation is crucial for their optoelectronic applications, yet difficult to rationalize due to the interplay of enthalpic, configurational, and dynamical effects. Here we present a simple thermodynamic framework for multicomponent halide perovskites of composition FA$_{1-x}$MA$_{x-y}$Cs$_y$Pb(I$_{1-z}$Br$_z$)$_3$, based on \textit{ab initio} molecular dynamics. By decomposing the free energy of mixing into enthalpic, configurational, and rotational entropic contributions, we show that although the enthalpy of mixing is generally positive, the solid solutions are thermodynamically stable against phase separation due to the large configurational entropy associated with random substitution on cation and halide sublattices. Mixing reduces the rotational entropy of the organic cations, partially offsetting the configurational stabilization. However, within our model, this rotational penalty is not sufficient to overcome the configurational driving force, and a curvature analysis within a regular-solution model does not predict a miscibility gap for any of the mixing channels considered. Analysis of hydrogen-bond dynamics shows that MA--Y (Y = I, Br) interactions are more persistent than FA--Y interactions, while the dominant FA-donated \hb{N}{I} hydrogen bonds remain nearly composition-invariant. Cs-containing mixtures, in which Cs$^{+}$ forms no hydrogen bonds, can nevertheless be thermodynamically stable. These results demonstrate that hydrogen bonding does not control thermodynamic stability in mixed halide perovskites. Instead, phase stability is governed by the balance between strong configurational entropy and a smaller, systematically destabilizing rotational-entropy correction.
\end{abstract}


Hybrid organic-inorganic halide perovskites (HOIHP) have attracted intense interest due to their wide range of optoelectronic applications, including in photovoltaics \cite{reviewperovskite2016,review_tandempsc2020}, light-emission devices \cite{review_peLEDS2021}, photodetectors \cite{review_detector_psk2022,review_detector_psk}, field-effect transistors \cite{review:hoipfet}, and photocatalysts \cite{review_psk_photocal2020,review_psk_photocal2022}. Perovskite solar cells (PSCs), employing HOIHP as light absorbers, have reached certified power conversion efficiencies (PCEs) of 27\% \cite{nrelchart}, and are promising candidates for tandem silicon-perovskite architectures.

\begin{figure}[!]
    \centering  \includegraphics[width=4cm]{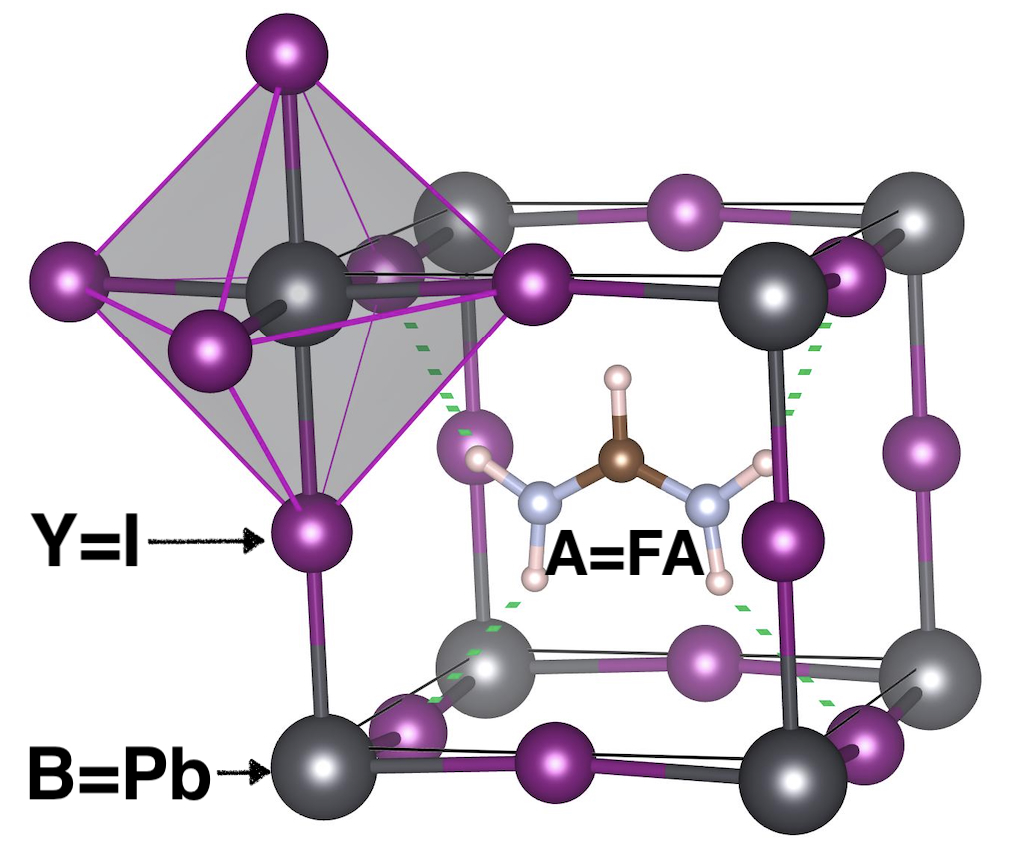}
    \caption{Unit cell of cubic perovskite \FApbi{}. Iodide anions outside unit cell are added at top left corner to show a PbI$_6$ octahedron. Green dotted lines indicate hydrogen bonds.}
    \label{fig:psk}
\end{figure}

The perovskite structure, derived from CaTiO$_3$, is described by the general formula ABY$_3$, where the B-site is surrounded by a three-dimensional network of corner-sharing Y$_6$ octahedra, and the A-site cation occupies the cavities created by this framework (Figure \ref{fig:psk}). In HOIHP, the size of the A-site cation is constrained by the Goldschmidt tolerance factor (0.8–1.0) \cite{Lee2022}, which limits the most commonly used organic cations to methylammonium (MA=\ma{}) and formamidinium (FA=\fa{}), while larger cations lead to lower-dimensional structures \cite{papavassiliou1992,mitzi1996-2Dpsk,mitzijacs1995,grancini2019}.

Early PSCs were based on \MApbi{} and \FApbi{} \cite{kojima2009,firstfapicell}; however, \MApbi{} is prone to decomposition and \FApbi{} undergoes detrimental phase transitions at room temperature. Significant improvements in stability and performance have therefore been achieved through alloying \cite{review_mixes_ono2017,record23.4psc,record23.3psc,record_25.2psc}. In mixed-cation HOIHP, the A-site is occupied by combinations of FA, MA, and/or Cs, while the B- and Y-sites may also be substituted to reduce toxicity (e.g. Pb/Sn) or tune the band gap and stability (I/Br/Cl). In particular, A-site alloying provides a means to fine-tune the lattice parameters and mitigate interfacial strain in devices.

Hydrogen bonds (HBs) can form in HOIHP between the hydrogens of the organic cations and the surrounding halide anions (Figure \ref{fig:psk}). According to IUPAC definition\cite{Arunan2011}, ``the hydrogen bond is an attractive interaction between a hydrogen atom from a molecule or a molecular fragment X$-$H in which X is more electronegative than H, and an atom or a group of atoms in the same or a different molecule, in which there is evidence of bond formation. In HOIHP, two main types of HBs can occur\cite{Varadwaj2019,Garrote2023}: \hb{N}{Y} and \hb{C}{Y}, where Y=Br, Cl, or I. These interactions have been proposed to influence structural stability\cite{HB_cheetham2015}, ion migration\cite{Zhang2022_migration}, and vibrational properties.

However, HBs in HOIHP are unstable, breaking and reforming on picosecond timescales\cite{saleh} even at temperatures as low as 50~K\cite{Garrote2024}, where the organic cations do not rotate. The associated activation energies in \MApbbr{} are small and comparable to thermal fluctuations \cite{Garrote2024,Svane2017}, suggesting that individual HBs are weak and transient. While these interactions provide important insight into the local coupling between organic and inorganic sublattices, their actual role in determining the thermodynamic stability of mixed HOIHP remains unclear.

Previous computational studies of FA-Cs systems have shown that Cs incorporation can induce coherent octahedral tilting, slightly restrict FA rotations, and modify non-covalent interactions within the lattice.\cite{Ghosh2017,Ghosh2018} These effects have been associated with enhanced structural stability. However, most available studies on mixed A-site halide perovskites primarily focus on structural, electronic, or phase-stability aspects.\cite{Dalpian2019,Unlu2020,Francisco-Lopez2020} While thermodynamic arguments are often invoked, they are generally inferred indirectly rather than obtained from an explicit decomposition of the mixing enthalpy, configurational entropy, and hydrogen-bond dynamics.

It is therefore not clear from previous studies what the role of hydrogen bonds is in the stabilization of hybrid perovskite solid solutions. In this work, we address this open question by performing a combined thermodynamic, structural, and dynamical study of A-site and A+Y-site mixed halide perovskites. Using ab initio molecular dynamics (AIMD), we compute the enthalpy ($\Delta H_\text{mix}$), configurational and rotational entropy ($\Delta S_\text{mix}^\text{conf}$, $\Delta S_\text{mix}^\text{rot}$), and effective Gibbs free energy ($\Delta G_\text{mix}^\text{eff}$) of mixing at 350~K for a series of pure and mixed compositions, including FA/MA and FA/Cs alloys at a 7/8 \FApbi{} ratio with simultaneous halide substitution. In parallel, we analyze the lifetimes and rebonding probabilities of \hb{N}{Y} HBs using time correlation functions, focusing on the effects of system composition.

In our simulations, all mixed perovskites were modeled in the cubic perovskite framework. We adopt a 4$\times$4$\times$4 supercell size, which was used and validated in our previous AIMD study of (FAPbI$_3$)$_{1-x}$(MAPbBr$_3$)$_{x}$, where $x$=1/8, containing 64 A sites, 64 Pb sites, and 192 halide sites.\cite{jmca2022} Atomic-scale disorder in the mixed configurations was generated using a special quasirandom structure (SQS)-based protocol as in Ref.~\citenum{jmca2022}, selecting configurations that reproduce near-random short-range correlations on the mixed sublattices and near-binomial local-environment statistics. Molecular A-site cations (when present) were then explicitly introduced with randomized orientations. Using the same protocol and supercell size, here we generated three additional mixed perovkites at the mixing fraction $1/8$, differing by which sublattices are disordered: (\textit{i}) A mix (cation-only): \fapimapi{}, (\textit{ii}) Y mix (halide-only): \fapifapb{}, and (\textit{iii}) A(Cs)+Y mix (cation-halide): \fapicspb{}. In all cases, the Pb sublattice is fully occupied, and the SQS optimization targets only the sublattices that are compositionally mixed (A, Y, or both), ensuring comparable cell size and sampling across systems.

AIMD simulations were carried out with the same DFT/MD setup (thermostatting protocol, timestep, basis sets, pseudopotentials, and dispersion treatment) as in Ref.~\citenum{jmca2022}. In addition to the mixed supercells described above, we also simulated the corresponding pure end members for each mixing family to provide reference baselines. All trajectories were generated under NVT conditions at $T=350$~K, and each system was propagated for 18~ps of production dynamics after thermalisation. These equilibrated production trajectories are used as input for the thermodynamic and hydrogen-bond analyses reported below.

The enthalpy of mixing per formula unit was obtained from equilibrium AIMD trajectories as
\begin{equation}
  \Delta H_\text{mix}(x)
  = \big\langle U_\text{mix}(x) \big\rangle 
   - (1-x)\langle U_A \rangle - x \langle U_B \rangle ,
\end{equation}
where $\langle U\rangle$ denotes a time average of the total energy per formula unit over the production portion of each trajectory (same $T$ and ensemble for mixed and reference systems). Uncertainties were estimated from the energy time series by accounting for equilibration and time correlation, using an effective number of independent samples $N_\text{eff}$ to compute the standard error of the mean, \(\mathrm{SEM}(U)=\sigma/\sqrt{N_\text{eff}}\), $\sigma$ being the standard deviation. The uncertainty in $\Delta H_\text{mix}$ was then obtained by standard error propagation from the SEMs of the mixed and reference energies and is reported as one standard deviation.

The configurational entropy of mixing was evaluated within the ideal solution approximation, following the same assumption as in Ref.~\citenum{jmca2022}. For binary mixing on a given sublattice (A or Y), the contribution per formula unit is
\begin{equation}
\Delta S_\text{mix}^\text{conf}(x) = - n_s k_\mathrm{B} \Big[x \ln x + (1-x)\ln(1-x)\Big],
\end{equation}
where $n_s$ is the number of mixed sites per ABY$_3$ formula unit. Thus, 
$n_s=n_A=1$ for A-mixing, $n_s=n_Y=3$ for Y-mixing, and $n_s=n_A+n_Y=4$ 
for simultaneous A+Y mixing, consistent with the expression used in Ref.~\citenum{jmca2022}. 

Because molecular A-site cations (FA and MA) can rotate within the inorganic cage, the total Gibbs free energy of the mixed perovskites includes an orientational entropy contribution and can be derived as \(\Delta G_\text{mix}^\text{tot} = \Delta H_\text{mix} - T(\Delta S_\text{mix}^\text{conf} + \Delta S_\text{mix}^\text{rot})\) per formula unit. When two distinct cations mix, their reorientational dynamics and thus their rotational entropies may change.  The rotational entropy of mixing,  \(\Delta S_\text{mix}^\text{rot}\), measures the difference between the orientational entropy of the mixture and the compositionally--weighted entropies of the pure end members. 
\(\Delta S_\text{mix}^\text{rot}\)
can be evaluated as a species--weighted sum of estimated rotational entropy changes.
\begin{equation}
\Delta S_\text{mix}^\text{rot}(x)=\sum_{\alpha} x_{\alpha}\,\Delta S_\mathrm{rot}^{\alpha},
\quad
\Delta S_\mathrm{rot}^{\alpha} \approx -\gamma\,k_\mathrm{B}\ln\!\left(\frac{\tau_{\alpha}^{\mathrm{mix}}}{\tau_{\alpha}^{\mathrm{pure}}}\right),
\label{eq:rotmix-methods}
\end{equation}
where \(x_\alpha\) is the molar fraction of species \(\alpha\) on the A sublattice, and \(\tau_{\alpha}\) is the corresponding rotational correlation time. The non-dimensional factor $\gamma = 0.185$ is obtained from the classical hindered-rotor model with a cosine potential,\cite{pitzer1942energy} by linearising the entropy-loss as a function of the rotational barrier. This approach allows us to estimate the rotational contribution to the mixing entropy without the long simulations required for an explicit rotational entropy calculation (from the distribution of orientations of each cation, which is difficult to converge). Eq. \ref{eq:rotmix-methods} implies that when there is no change in the rotational correlation times between the pure systems and the mixtures, there is no rotational contribution to the mixing entropy. But if, as it happens here, the cation rotation is more hindered in the solid solution than in the pure compounds, the correlation time will be longer in the former, and the contribution to the mixing entropy will be negative. The full derivation of Eq.~\ref{eq:rotmix-methods} and its assumptions are provided in the Supporting Information (SI). The correlation times were obtained from normalized orientational autocorrelation functions \(C(t)\) of the main A-molecular axes computed from the AIMD trajectories using the TRAVIS code\cite{travis}. The \(C(t)\) definition, together with details of our operational determination of the correlation time as the orientational half-time \(\tau\equiv t_{1/2}\) (defined as the first time such that \(C(t_{1/2})=0.5\)), are also provided in the SI.

The statistics of HBs has been collected from MD trajectories using TRAVIS as well. In particular, we computed HB populations and time-correlation descriptors, including combined distribution functions (CDF) and aggregate correlation functions (ACF), to characterize HB geometries and lifetimes in the mixed and pure systems. 
The CDFs, shown in Figure S4-S5 in Supporting Information, are two-dimensional histograms of the frequency of a pair of parameters. In the CDFs here presented, that pair of parameters comprises the HB H$-$Y distance and the $\N-\mathrm{H}-\Y$ angle. 
The ACF gives the probability that a link between two molecules, e.g., MA and Br linked by a HB, persists during time $t$. The continuous ACF can be cast as
\begin{equation}
C_{C}^{HB}(t)=\frac{A}{N_1 N_2} \sum_{i=1}^{N_1}\sum_{j=2}^{N_2}\left\langle\beta_{ij}(t^{\prime})\widetilde\beta_{ij}(t^{\prime}+t) \right\rangle_{t^{\prime}},
\label{ec:cdacf}
\end{equation}
where $i$ and $j$ run across all molecules of a given species, $i$ for an organic cation and $j$ for the halogen, respectively. The function 
$\beta_{ij}(t^{\prime})=1$ if the link exists at time $t'$ or 0 if not. The function $\widetilde\beta_{ij}(t^{\prime}+t)=1$ if the link exists during 
the whole time interval $(t',t'+t)$. 
$\langle ...\rangle_{t'}$ represents the average across the time instants $t'$ in the MD simulation. Finally, $A$ is a normalization factor to accomplish $C_{C}^{HB}(0)=1$. 
The lifetime of the aggregate is given by the integral\cite{travis} 
\begin{equation}
    \tau=2\int_{0}^{\infty}C_{C}^{HB}(t)dt .
\end{equation}
The intermittent ACF $C_{I}^{HB}(t)$ is 
defined by an expression simular to Eq.~\ref{ec:cdacf} with the function 
$\widetilde\beta_{ij}$ replaced by $\beta_{ij}$. Thus, $C_{I}^{HB}(t)$ gives 
the probability of the link existence 
if it was established a time $t$ in the past, not requiring the continuous existence 
at intermediate times. 

We now present the calculated thermodynamic trends and hydrogen-bond dynamics.

\begin{table*}[th]
\setlength{\tabcolsep}{6pt} 
\renewcommand{\arraystretch}{1.4}
\centering
\begin{tabular}{
  @{} l                
  l                    
  S[table-format = +1.2]         
  S[table-format = 1.2e1]        
  S[table-format = 1.2e1]        
  S[table-format = -1.2]         
  @{}}
\toprule
\textbf{Mixing site} & \textbf{Perovskite} &
  {$\Delta H_\text{mix}$} &
  {$\Delta S_\text{mix}^\text{conf}$} &
  {$\Delta S_\text{mix}^\text{rot}$} &
  {$\Delta G_\text{mix}^\text{tot}$} \\
& & 
\multicolumn{1}{c}{(kJ mol\(^{-1}\))} &
\multicolumn{1}{c}{(kJ K\(^{-1}\) mol\(^{-1}\))} &
\multicolumn{1}{c}{(kJ K\(^{-1}\) mol\(^{-1}\))} &
\multicolumn{1}{c}{(kJ mol\(^{-1}\))} \\
\midrule
A       & \fapimapi{}  & \multicolumn{1}{c}{$-0.11(21)$} & 3.13e-3 & \multicolumn{1}{c}{$-5.35\times10^{-4}$}   & \multicolumn{1}{c}{$-1.02(21)$} \\   
Y       & \fapifapb{}  & \multicolumn{1}{c}{$+0.70(24)$} & 9.40e-3 & \multicolumn{1}{c}{$-4.01\times10^{-4}$} & \multicolumn{1}{c}{$-2.45(24)$} \\
A + Y   & \fapimapb{}  & \multicolumn{1}{c}{$+0.91(22)$} & 1.25e-2 & \multicolumn{1}{c}{$-4.48\times10^{-4}$}   & \multicolumn{1}{c}{$-3.32(22)$} \\
A + Y   & \fapicspb{}  & \multicolumn{1}{c}{$+1.06(19)$} & 1.25e-2 & \multicolumn{1}{c}{$-9.89\times10^{-4}$}   & \multicolumn{1}{c}{$-2.98(19)$} \\
\bottomrule
\end{tabular}
\caption{
Thermodynamic potentials of mixing (per formula unit) for the studied perovskites (\(ABY_{3}\)) at 350~K. The configurational entropy corresponds to the ideal mixing of
\(N\) substitutional sites (A and/or Y), while \(\Delta S_\text{mix}^\text{rot}\) is the loss of rotational entropy estimated from molecular dynamics trajectories (see the Supporting Information (SI) for the derivation and assumptions).
The total free energy is \(\Delta G_\text{mix}^\text{tot} = \Delta H_\text{mix} - T(\Delta S_\text{mix}^\text{conf} + \Delta S_\text{mix}^\text{rot})\).
Uncertainties correspond to one standard deviation.}
\label{tab:thermo}
\end{table*}

\begin{table*}[t]
\footnotesize                                 
\setlength{\tabcolsep}{2pt}                   
\renewcommand{\arraystretch}{1.5}             

\centering                                    
\begin{adjustbox}{width=\textwidth}           
\begin{tabularx}{\textwidth}{@{} l
  S S S  S S S  S S S  S S S @{}}
\toprule
\textbf{Perovskite} &
\multicolumn{3}{c}{\textbf{(FA) N–H···I}\cite{noteHB}} &
\multicolumn{3}{c}{\textbf{(FA) N–H···Br}} &
\multicolumn{3}{c}{\textbf{(MA) N–H···I}} &
\multicolumn{3}{c}{\textbf{(MA) N–H···Br}} \\[-0.4em]
 & {$PD$$^a$} & {$\tau$\,(ps)} & {$N_n$}
 & {$PD$$^a$} & {$\tau$\,(ps)} & {$N_n$}
 & {$PD$$^a$} & {$\tau$\,(ps)} & {$N_n$}
 & {$PD$$^a$} & {$\tau$\,(ps)} & {$N_n$} \\
\midrule
\fapi{}       & 0.223 & 0.156 & 2.0 & {--} & {--} & {--} & {--} & {--} & {--} & {--} & {--} & {--} \\
\fapb{}       & {--}   & {--}   & {--} & 0.298 & 0.197 & 2.6 & {--} & {--} & {--} & {--} & {--} & {--} \\
\mapi{}       & {--}   & {--}   & {--} & {--} & {--} & {--} & 0.271 & 0.171 & 1.7 & {--} & {--} & {--} \\
\mapb{}       & {--}   & {--}   & {--} & {--} & {--} & {--} & {--} & {--} & {--} & 0.312 & 0.229 & 2.2 \\
\fapimapi{}   & 0.234 & 0.156 & 2.0 & {--} & {--} & {--} & 0.286 & 0.193 & 1.8 & {--} & {--} & {--} \\
\fapifapb{}   & 0.233 & 0.158 & 1.7 & 0.302 & 0.213 & 0.4 & {--} & {--} & {--} & {--} & {--} & {--} \\
\fapimapb{}   & 0.238 & 0.161 & 1.7 & 0.325 & 0.242 & 0.4 & 0.297 & 0.183 & 2.0 & 0.457 & 0.275 & 0.2 \\
\fapicspb{}   & 0.233 & 0.151 & 1.7 & 0.312 & 0.226 & 0.4 & {--} & {--} & {--} & {--} & {--} & {--} \\
\bottomrule
\end{tabularx}
\end{adjustbox}

\caption{Hydrogen-bond lifetimes \(\tau\) between A-site cations and Y-site anions in pure and mixed perovskites, obtained from continuous time correlation functions at 350~K, \(PD\) measures bond reformation probability, and \(N_n\) is the average number of bonded neighbours.}
\label{tab:lifetimes}
\raggedright
$^a$ \(PD = C_\mathrm{int}(t=\tau) - C_\mathrm{cont}(t=\tau)\), where \(C_\mathrm{int}\) and \(C_\mathrm{cont}\) are the intermittent and continuous correlation functions, respectively.
\end{table*}

\begin{figure}
    \centering
    \includegraphics[width=1.02\linewidth]{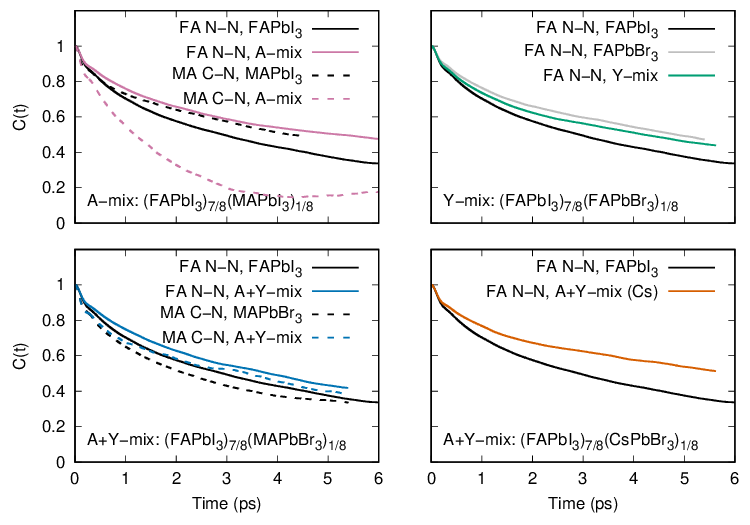}
    \caption{Orientational autocorrelation functions $C(t)$ for the FA N--N axis and the MA C--N axis in pure and mixed perovskites (A-site, Y-site, and A+Y mixing). Slower decay (larger correlation time) indicates slower reorientation; correlation times are obtained from fits as described in the SI.}

    \label{fig:cation_reorientation}
\end{figure}

In previous work on mixed perovskites, we reported that cation reorientation dynamics in \fapimapb{} differ from those of the corresponding pure compounds, although the observed effect was relatively modest.\cite{jmca2022} In the present study, by extending the analysis to additional mixing channels and compositions, we find that changes in molecular reorientation dynamics are significantly more pronounced and strongly dependent on the nature of the mixed sublattices. In particular, A-site mixing and the Cs-containing A+Y mixture exhibit substantial modifications of cation rotation, while more moderate effects persist in Y-site mixing and in \fapimapb{}. These observations motivate a quantitative assessment of the role of molecular rotation in the thermodynamics of mixing. To this end, we analyze the reorientation dynamics of representative molecular vectors obtained from molecular dynamics trajectories (Figure \ref{fig:cation_reorientation}) and use the resulting rotational correlation times to estimate the rotational entropy change upon mixing.

The reorientation dynamics of the FA N--N vector and the MA C--N vector (Figure \ref{fig:cation_reorientation}), reveal a composition-dependent modification of molecular rotation. In all mixed compositions, the reorientation of FA is systematically slowed down relative to pure \fapi{}, indicating a restriction of its accessible orientational phase space upon mixing. This systematic slowdown indicates that both A-site substitution (FA/MA or FA/Cs) and halide substitution (I/Br) perturb the cage in a way that reduces FA rotational freedom, pointing to a mixing-induced orientational constraint on FA that contributes negatively to \(\Delta S_\text{mix}^\text{rot}\). In contrast, MA reorientation exhibits a mixing dependence: it is markedly accelerated in \fapimapi{}, whereas it is slowed in \fapimapb{} relative to \mapi{} and \mapb{}.
In line with the above trends, solid-state NMR relaxometry indicates that cation reorientation is strongly affected by the inorganic-lattice symmetry: FA slows near the cubic-tetragonal transition (and may slow down further in mixtures with Cs), while MA shows a $\sim$2$\times$ faster rotation at room temperature in cubic FA$_x$MA$_{1-x}$PbI$_3$ compared with MAPbI$_3$.\cite{Mishra2023}

Based on these reorientation dynamics, we estimate the rotational entropy change upon mixing, $\Delta S_\text{mix}^\text{rot}$, providing a microscopic basis for the rotational entropy term included in the thermodynamic analysis. The underlying reorientation times $\tau_\alpha$ (FA N--N and MA C--N), taken in practice as the fit-free half-time \(t_{1/2}\) from the  AIMD autocorrelation functions, are reported in Table~S1 of the SI. Table~\ref{tab:thermo} summarizes the thermodynamic quantities for all mixed perovskites at 350~K, including enthalpies of mixing, the ideal configurational entropy, the rotational entropy change extracted from molecular-dynamics trajectories, and the corresponding total Gibbs free energies \(\Delta G_\text{mix}^\text{tot}\).

A direct comparison of the entropy contributions in Table~\ref{tab:thermo} reveals a clear separation of roles between configurational and rotational entropy. The configurational entropy of mixing is positive and constitutes the dominant stabilizing contribution for all compositions, increasing systematically with the number of mixed sites. In contrast, the rotational entropy is negative for all systems: because $\frac{\tau^{\mathrm{mix}}}{\tau^{\mathrm{pure}}} > 1$, it acts as a destabilizing contribution that partially offsets the configurational entropy.

\begin{figure}
    \centering
    \includegraphics[width=1.02\linewidth]{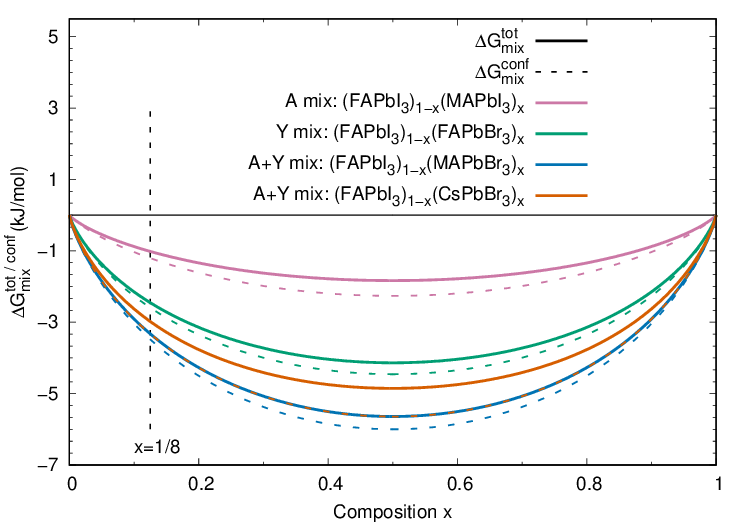}
    \caption{Thermodynamic bowing curves for $\Delta G_\text{mix}^\text{tot}$ (solid lines) and $\Delta G_\text{mix}^\text{conf}$ (dashed lines) at 350~K for A-site, Y-site, and A+Y-site mixed perovskites. The vertical dashed line marks the simulated composition $x=1/8$.}
    \label{fig:thermo-bowing}
\end{figure}

To assess the stability of the solid solutions against phase separation, it is not enough to consider the sign of the mixing free energy at a single composition $x$: one must consider the curvature of the mixing free energy as a function of $x$. Extending the AIMD-based thermodynamic analysis to a grid of $x$ values is computationally too expensive, but we can analyze the curvature of the mixing Gibbs free energy within the framework of a regular-solution model, parameterized using the computed thermodynamic quantities at $x=1/8$ (Figure~\ref{fig:thermo-bowing}).
Here $W$ is an enthalpic parameter obtained from the bowing of $\Delta H_\text{mix} = W x(1-x)$, while $\lambda$ captures the rotational-entropy correction via $-T\Delta S_\text{mix}^\text{rot} = T\lambda x(1-x)$. With this definition, $\lambda>0$ corresponds to a loss of rotational entropy upon mixing.
In this model, a miscibility gap occurs when $W_\mathrm{eff}>2n_{s}k_\mathrm{B}T$, where $W_\mathrm{eff}=W+T\lambda$ and $n_s$ is the number of mixed sites per formula unit. This criterion diagnoses the enthalpy/entropy tendency toward phase separation within the regular-solution approximation.

Although $\Delta G_\text{mix}^\text{tot}$ is negative at the simulated composition for all systems, the propensity for macroscopic phase separation is governed by the curvature of $\Delta G_\text{mix}(x)$. Quantitatively, for all mixing channels considered here the curvature criterion does not predict a miscibility gap at 350~K. Specifically, we obtain $W_\mathrm{eff}=7.67$~kJ~mol$^{-1}$ for \fapifapb{} (Y-mix), $0.74$~kJ~mol$^{-1}$ for \fapimapi{} (A-mix), $9.72$~kJ~mol$^{-1}$ for \fapimapb{} (A+Y mix), and $12.88$~kJ~mol$^{-1}$ for the Cs-containing A+Y mixture \fapicspb{}. All values remain below the corresponding thresholds, indicating that within this regular-solution analysis, the free-energy curves remain convex and the solid solutions are thermodynamically stable against macroscopic phase separation.

These trends are consistent with previous atomistic and experimental studies suggesting that mixed-cation and mixed-halide perovskites can form stable solid solutions over broad composition ranges, with configurational entropy and strain accommodation playing central roles in the thermodynamic balance.\cite{Yi2016,Dalpian2019} In the specific case of FA/Cs mixtures, Cs incorporation has been reported to stabilize the perovskite phase by reducing local lattice frustration and promoting a more coherent octahedral tilting pattern.\cite{Ghosh2018}
This effect agrees with FA/MA mixing, in which MA is often associated with increased dynamic disorder and volatility rather than with the suppression of structural instabilities. In fact, temperature-dependent experiments report FA/MA solid solutions across the full composition range and a cubic phase at high temperature for all $x$.\cite{Francisco-Lopez2020}

In our simulations, Y-mixing and A+Y mixing are strongly stabilized by configurational entropy, whereas FA/MA A-site mixing shows a smaller but still negative $\Delta G_\text{mix}^\text{tot}$ at this composition. The Cs-containing A+Y mixture exhibits the largest rotational-entropy penalty, consistent with more constrained FA dynamics, yet remains thermodynamically favorable and well below the miscibility threshold in the curvature analysis.
Overall, these results support a picture in which configurational entropy provides the dominant stabilization, while rotational entropy contributes a systematic, typically destabilizing correction that can modulate the curvature of $\Delta G_\text{mix}(x)$.

Notably, the \fapimapb{} (A+Y mixed) composition exhibits the most favorable (most negative) $\Delta G_\text{mix}(x)$ among the systems studied, consistent with the widespread use of mixed-cation/mixed-halide absorbers in high-performance perovskite solar cells.\cite{record23.3psc,record23.4psc,record_25.2psc} Mixed FA/MA and I/Br formulations form the basis of many state-of-the-art device recipes, where compositional engineering is commonly rationalized in terms of improved phase purity and enhanced thermodynamic stability driven primarily by configurational entropy rather than strongly exothermic interactions.\cite{Saliba2016,Choe2021,Chen2025} Our results provide an equilibrium thermodynamic counterpart to this empirical success: simultaneous A- and Y-sublattice disorder maximizes the entropic stabilization and yields a strongly favorable mixing free energy. We emphasize, however, that device record efficiencies also depend critically on crystallization pathways, defect chemistry, and interfaces, and mixed-halide systems may still exhibit light-induced segregation under operating conditions.\cite{Choe2021}

Having established that all mixing channels considered here are thermodynamically stable at 350~K (with no miscibility gap predicted within our regular-solution analysis), we next examine whether hydrogen bonding provides a microscopic origin for the observed differences among mixing channels. We address this point by analyzing the lifetimes and rebonding probabilities of the hydrogen bonds.

\begin{figure}[!]
  \centering
  \includegraphics[width=0.47\textwidth]{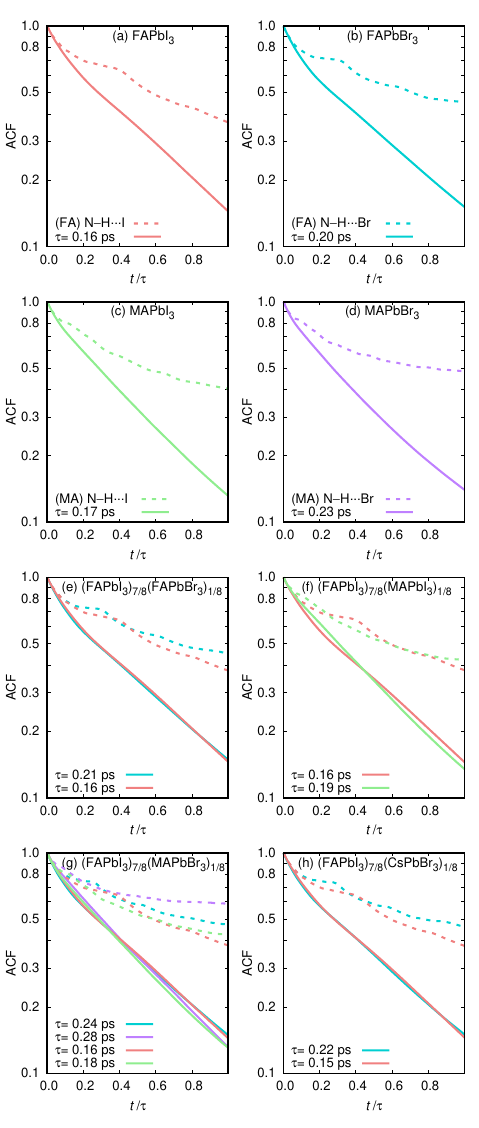}
\caption{\textbf{(a)-(h)} ACFs of \hb{N}{I} and \hb{N}{Br} HBs calculated as continuous (continuous line) and intermittent types (dashed line) of the pure and mixed compounds.}
\label{fig_dacf}
\end{figure}

We consider that a \hb{N}{Y} HB (Y=Br, I) is established when two geometrical conditions are simultaneously fulfilled:(1) the distance $d(H-Y)\le 3$~\AA{}, and 
(2) the angle $\measuredangle (\N-\mathrm{H}-\Y)$ is between $135^{\circ}$ and $180^{\circ}$. This procedure was established\cite{Garrote2023} by analyzing the CDF of $d(H-Y)$ and $\measuredangle(\N-\mathrm{H}-\Y)$, for the cases of \fapi{}, \mapb{}, and \fapimapb{}. That analysis was complemented by calculations of the reduced density gradient of the electronic density. 
Figure S4-5 in Support Information shows the CDFs for the former compounds and the new compounds here studied. It can be seen that the aforementioned conditions enclose a region of high 
probability for all the studied materials. Moreover, for each particular HB, e.g., (FA)\hb{N}{Br}\cite{noteHB}, the appearance of the CDF is the same for all compounds.
Hence, we keep 
these geometrical criteria to determine the dynamical properties of the HBs in these materials. 

Figure \ref{fig_dacf} shows the continuous and intermittent ACF for all the studied materials. 
The functions in the graphs are displayed up the HB lifetime ($t/\tau=1$). 
The HB lifetimes $\tau$ extracted from the continuous ACF are summarized in Table~\ref{tab:lifetimes}. 
The intermittent ACF cannot decay to zero because the confinement of the FA and MA cations implies that all 
broken HBs eventually reform. Nevertheless, the difference between the intermittent and continuous ACF is a measure of the fraction of cation-halogen pairs that rebond after HB breaking instead of evolving to form a HB with different pairs. In particular, the difference 
between both ACFS at the lifetime is reported as probability difference ($PD$) in Table~\ref{tab:lifetimes}. Overall, HB lifetimes remain in a narrow sub-ps range across pure and mixed systems, whereas $PD$ and the coordination number $N_n$ mainly reflect local halide availability (notably the low Br fraction in the mixed-halide cases). This indicates that hydrogen bonds adapt to the mixed environment, but do not provide a direct thermodynamic driving force for the distinct mixing stabilities discussed above.

The following trends are observed. The (FA)\hb{N}{I} HB has almost the same lifetime ($\pm$ 3$-$4 \%) and $PD$ for all compositions. This is probably due to the fact that all compounds has either 100\% or 87.5\% FA and I content. The number of neighbors (halogen neighbors of FA or FA neighbors of I) is proportional to the iodine content. The number of neighbors is the average number of halogen anions hydrogen-bonded to each FA or MA cation. 
More significant variations can be observed for the HB bonds that become minoritary in mixed compounds. The (FA)\hb{N}{Br} bonds in the pure 
compound \fapb{} show longer lifetime and greater number of neighbors than the (FA)\hb{N}{I} in pure \fapi{}. In the mixed compounds the number of neighbors decreases strongly because Br content becomes minoritary. The decrease in the number of neighbors correlates with an increase of lifetime and probability difference. In other words, there are fewer bonds, but once established they last longer. This behaviour is characteristic of minority–species effects, where reduced local configurational freedom restricts bond exchange and leads to enhanced persistence of the remaining HBs. The same behaviour can be appreciated for the (MA)\hb{N}{Br} HBs present in \mapb{} and \fapimapb{}.

Overall, the HBs statistics reveal a consistent hierarchy in both lifetime and rebonding probability: Br$-$based interactions are more persistent and prone to rebonding than their iodide counterparts, whereas MA$-$centered bonds display greater orientational memory than those involving FA. Conversely, (FA)\hb{N}{I} HB remain essentially unaffected by the composition, underscoring the intrinsic character of this interaction within the PbI{} lattice.

The FA/Cs mixture (\fapicspb{}) deserves special attention. Its (FA)\hb{N}{I} lifetimes are slightly shorter than in \fapi{}, but $PD$ is marginally higher while $N_n$ decreases only modestly. Thus, FA-I interactions in \fapicspb{} break on similar timescales but tend to reform with the same acceptor, indicating a confined and persistent FA-halide environment rather than a weaker HB network. This is consistent with previous ab initio studies reporting restricted cation-lattice dynamics and shorter \hb{N}{I} distances upon Cs incorporation.\cite{Ghosh2017,Ghosh2018} Importantly, these dynamical differences do not systematically track the mixing thermodynamics (Table~\ref{tab:thermo}) and do not provide a microscopic explanation for the overall stability of \fapicspb{}, reinforcing that hydrogen bonding is not the primary driver of thermodynamic stability.

In the \fapimapb{} solid solution, MA forms more persistent \hb{N}{Y} interactions than FA (Table~\ref{tab:lifetimes}), particularly for Br where $\tau$ and $PD$ are the largest ones despite the low coordination $N_n$.
Thus, the mixed-cation environment supports coexisting HB regimes with distinct persistence and rebonding behavior (FA--I versus MA--I/Br), which can promote local dynamical heterogeneity even though the dominant FA--I lifetime itself remains nearly composition-invariant. This heterogeneity may contribute to local frustration, although it does not map directly onto $\Delta H_\text{mix}$ or $\Delta G_\text{mix}^\text{tot}$.

Reorientational A-site dynamics (Figure S2, SI) further support this picture. FA reorientation slows systematically in mixed compositions, whereas MA reorientation is mixing-channel dependent (slower in \fapimapi{} but faster in \fapimapb{} relative to the corresponding pure compound). Since (FA)\hb{N}{I} HB lifetimes remain nearly unchanged, the slowdown reflects increased anisotropy and heterogeneity of the local environment rather than stronger individual interactions. The corresponding loss of rotational entropy is small but consistent with the computed \(\Delta S_\text{mix}^\text{rot}\) values. As shown before, these dynamical contributions are secondary to the dominant configurational entropy term that stabilizes mixing.

To test whether HB behavior correlates with broader structural features, we compared HB lifetimes with the Root Mean Squared displacement (RMSD) across compositions (Figure S10 in SI). These global descriptors vary only modestly and do not correlate with either HB statistics or mixing enthalpies. A noteworthy exception is the FA/Cs system, where the FA center-of-mass RMSD closely follows that of Cs, indicating a high degree of dynamical compatibility that contrasts with the heterogeneity observed in FA/MA mixtures. These observations reinforce the conclusion that FA/MA and FA/Cs mixtures achieve comparable thermodynamic stability through distinct microscopic pathways, rather than through a dominant HB-based mechanism.

Taken together, these observations indicate that the hydrogen bond is not the primary thermodynamic driving force for mixing in the perovskite considered here.
The dominant FA--I HBs exhibit nearly invariant lifetimes and rebonding probabilities across compositions and therefore do not systematically track either $\Delta H_\text{mix}$, $\Delta G_\text{mix}^\text{tot}$, or the phase-stability trends inferred from free-energy curvature. Although Br-based and MA-centered interactions are typically longer-lived and can display enhanced rebonding (notably for MA--Br in \fapimapb{}), these local dynamical signatures do not translate into systematically more favorable mixing free energies. Conversely, Cs does not form hydrogen bonds, yet the Cs-containing mixture \fapicspb{} remains thermodynamically favorable at $x=1/8$ and does not exhibit a miscibility gap within our curvature analysis, indicating that hydrogen bonding is not a prerequisite for mixing stability in these systems.

In conclusion, we find that hydrogen bonds are not the primary thermodynamic driving force for the stability of mixed FA/MA/Cs halide perovskites. Although hydrogen bonding is essential to sustain the hybrid framework and modulate local interactions, their lifetimes and rebonding probabilities do not correlate with the calculated $\Delta H_\text{mix}$ or $\Delta G_\text{mix}$. Instead, stability is dominated by configurational entropy, with lattice strain accommodation, and a systematically destabilizing rotational-entropy correction providing secondary contributions.

Importantly, comparable thermodynamic stability can arise through different microscopic responses. In FA/MA solid solutions, the coexistence of two hydrogen-bonding cations promotes dynamical heterogeneity and local frustration of the HB network. In contrast, FA/Cs systems exhibit a more homogeneous dynamical response due to the absence of HB competition and the greater compatibility of the A-site species. In both cases, the HB network adapts to the mixed environment rather than dictating the thermodynamic outcome.

By combining thermodynamic analysis with HB statistics and cation reorientation dynamics, we show that HBs play a secondary, modulatory role in the stabilization of mixed halide perovskites. These materials instead achieve stability through a subtle interplay of configurational disorder, lattice response, and adaptive local interactions.

Note that our analysis is based on idealized atomistic models and equilibrium thermodynamics. In real materials, additional factors such as defects, strain gradients, illumination-induced segregation, and finite-size effects may influence phase stability. Within these limitations, our calculations indicate that all the solid solutions considered are thermodynamically stable against phase separation, without miscibility gaps at 350~K.

\begin{acknowledgement}
The authors acknowledge support from: Powered@NLHPC: This research/thesis was partially supported by the supercomputing infrastructure of the NLHPC (CCSS210001). Project supported by the Competition for Research Regular Projects, year 2023, code LPR23-04, Universidad Tecnológica Metropolitana. Project supported by the “Competition for Research Assistant Funding UTEM”, year 2025, code AI25-05 Universidad Tecnológica Metropolitana.
We are also grateful to the UK Materials and Molecular Modelling Hub for computational resources, which is partially funded by EPSRC (EP/P020194/1 and EP/T022213/1). EMP acknowledges travel support from Asociaci\'on Universitaria Iberoamericana de Postgrado.
\end{acknowledgement}


\begin{suppinfo}
The Supporting Information includes: (i) derivation of the rotational-entropy-of-mixing estimator for mixed-cation perovskites (including the operational definition of the reorientation time via the orientational half-time ($t_{1/2}$); (ii) additional cation reorientation dynamics results; (iii) estimation of uncertainties in the calculated thermodynamic functions; (iv) regular-solution model analysis of mixing free energies and phase stability; (v) detailed hydrogen-bond structural and time-correlation analyses; and (vi) root-mean-square displacement (RMSD) analysis. This material is available free of charge.
\textbf{Data Availability}: The raw AIMD trajectories and the processed datasets derived from them (TRAVIS outputs for hydrogen-bond analysis, cation reorientation dynamics, and mean-square displacement) are available on Zenodo at DOI: 10.5281/zenodo.18604579

\end{suppinfo}


\bibliography{refer}
\end{document}


\centering\section*{\fontsize{14}{15}\selectfont Supporting Information}
\title{Thermodynamic Stability and Hydrogen Bonds in Mixed Halide Perovskites}

\author{Liz Camayo-Gutierrez}
\author{Javiera Ubeda}
\author{Ana L. Montero-Alejo}
\email{amonteroa@utem.cl}
\affiliation
{Departamento de F\'isica, Facultad de Ciencias Naturales, Matem\'atica y del Medio Ambiente (FCNMM), Universidad Tecnol\'ogica Metropolitana, Jos\'e Pedro Alessandri 1242, \~Nu\~noa 7800002, Santiago, Chile}
\author{Ricardo Grau-Crespo}
\affiliation{School of Engineering and Materials Science, Queen Mary University of London, Mile End Road, London E1 4NS, UK}
\author{Eduardo Men\'endez-Proupin}
\email{emenendez@us.es}
\affiliation{Departamento de Física Aplicada I, Escuela Politécnica Superior, Universidad de Sevilla, Seville E-41011, Spain}

\maketitle
\newpage
\justify

\section{Rotational entropy of mixing for mixed-cation perovskites}

This section derives a practical expression for the rotational entropy of mixing, \(\Delta S_\text{mix}^\text{rot}\), in terms of species-specific rotational correlation times obtained from orientational autocorrelation functions.

The key idea is that any rotation-hindering potential barrier reduces the accessible phase space, lowering the entropy. One can estimate the entropy reduction from the effective energy barrier, which in turn can be estimated from the decay of the orientational auto-correlation function of the rotating cations, using relatively short molecular dynamics simulations.

To obtain an analytical expression for the orientational entropy loss, we map the reorientation of a given molecular axis into a one-dimensional (1D) rotor model in a periodic cosine potential,
\begin{equation}
V(\theta) = \frac{\Delta E}{2}\bigl(1 - \cos\theta\bigr),
\end{equation}
where \(\Delta E\) is the barrier height.  The relevant dimensionless parameter is \(y = \Delta E/(k_B T)\).  The orientational partition function and entropy difference relative to a free rotor can be evaluated exactly$^{1}$. The calculation involves modified Bessel functions \(I_0\) and \(I_1\):
\begin{equation}
\frac{\Delta S(y)}{k_B} = \ln I_0\!\left(\frac{y}{2}\right) - \frac{y}{2}\,\frac{I_1(y/2)}{I_0(y/2)},
\end{equation}
where \(\Delta S(y)\equiv S_{\text{orient}}(y) - S_{\text{orient}}^{\text{free}} \) is negative.  We define the positive entropy‐loss function
\begin{equation}
f(y) = -\Delta S(y) = k_B\left[ -\ln I_0\!\left(\frac{y}{2}\right) + \frac{y}{2}\frac{I_1(y/2)}{I_0(y/2)} \right].
\end{equation}

\begin{figure}[t]
    \centering
    \includegraphics[width=0.8\textwidth]{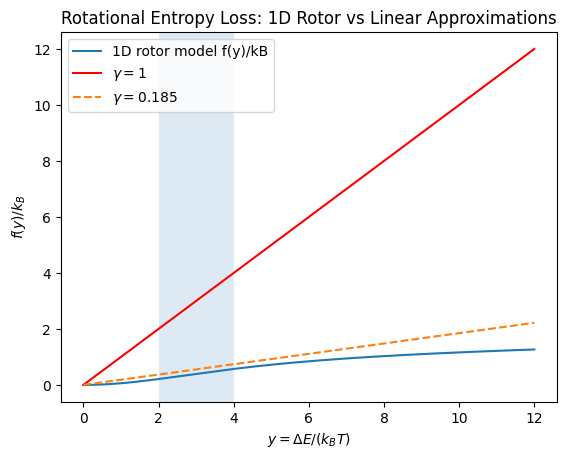}
    \caption{Normalized entropy loss \(f(y)/k_B\) for the 1D periodic hindered rotor.  The blue curve is the exact model [Eq.~(4)], the red line is the naive linear approximation \(y\), and the dashed black line is the best linear fit in the moderate barrier regime (\(2\le y\le4\)) with slope \(\gamma\approx0.185\).  The shaded region highlights the range of \(y\) relevant to organic cations in hybrid perovskites.}
    \label{fig:rotor}
\end{figure}

The function \(f(y)\) increases smoothly starting from zero (free rotation, \(y\to 0\)). The initial increase is very small, then becomes roughly linear, and finally saturates for very large barriers.  When the barriers are non-negligible but not too high either (up to a few times $k_BT$, near the shaded area in the figure), \(f(y)/k_B\) is well approximated as proportional to $y$, but with a slope much smaller than unity.  
\begin{equation}
f(y)/k_B \approx \gamma\,y,\qquad \gamma\approx 0.185, \label{eq:linear}
\end{equation}
where \(\gamma\) plays the role of an effective linear coefficient (much less than the naive value of unity).  Figure~\ref{fig:rotor} shows the exact \(f(y)/k_B\) alongside the naive line (slope~1) and the effective linear line with slope \(\gamma\approx 0.185\).

In an Arrhenius-like description of activated reorientation, the orientational correlation time \(\tau \) of a molecular axis is related to the barrier height by
\begin{equation}
\tau \approx \tau_0\,e^{\Delta E/(k_B T)} = \tau_0\,e^{y},\label{eq:arrhenius}
\end{equation}
where \(\tau_0\) is a prefactor determined by librational dynamics.  Taking the logarithm gives \(y = \ln(\tau/\tau_0)\), therefore the entropy loss of a single cation can be expressed as:

\begin{equation}
    \Delta S \approx -\gamma k_\mathrm{B} \ln(\tau /\tau_0) 
\end{equation}

We do not know $\tau_0$, but for two environments (e.g. a cation in the mixture and in a pure end member), we take the difference, so the linear approximation leads to a cancellation of $\tau_0$ and a simple expression for the rotational entropy change per cation:
\begin{equation}
\Delta S_{\mathrm{rot}}^{(\text{mix-pure})} \equiv -(f(y_{\text{mix}}) - f(y_{\text{pure}})) \approx -\gamma\,k_B\,\ln\!\left(\frac{\tau_{\text{mix}}}{\tau_{\text{pure}}}\right).\label{eq:deltaS}
\end{equation}
If, as happens in hybrid organic-inorganic perovskite solid solutions, mixing leads to additional hindering of rotation, the cation orientations will be more correlated over time and $\tau_{\text{mix}} > \tau_{\text{pure}}$, and therefore the rotational entropy of that molecular ion decreases upon mixing. Equation \eqref{eq:deltaS} is the final working formula used in our analysis.  It relates the change in rotational entropy to correlation times. 

The choice of a linear model with \(\gamma\approx0.185\) is, in summary, justified by the following arguments:

\begin{enumerate}
    \item The reorientational dynamics of the molecular axis are governed by an effective potential along one reaction coordinate (the slowest mode), even though the full orientational motion is two-dimensional (in $\theta$ and $\phi$).  The effective coordinate can be modelled by the 1D periodic potential.
    \item As seen above, the intermediate barrier regime relevant to room-temperature halide perovskites, the exact entropy loss function of this model is close to linear in \(y\) with a slope of \(\gamma \approx 0.185\).  
    \item Using ratios of correlation times eliminates the unknown prefactor \(\tau_0\) in this approximation. The linear coefficient \(\gamma\) thus becomes an effective proportionality constant mapping kinetic barriers (from \(\tau\)) to orientational entropy.
\end{enumerate}

In conclusion, a simple hindered-rotor model yields a closed-form expression for the rotational contribution to the entropy of mixing in a solid solution with rotating ions, that can be readily evaluated with parameters extracted from molecular dynamics simulations. In the moderate-barrier regime, the entropy loss is nearly linear in \(y=\Delta E/(k_B T)\) with an effective slope \(\gamma\approx0.185\).  Using the Arrhenius relation between barrier and correlation time, we derived Eq.~\eqref{eq:deltaS}, which expresses the change in rotational entropy upon mixing in terms of the ratio of relaxation times.  This approach naturally removes the unknown attempt time \(\tau_0\) and permits a simple estimation of the rotational contribution to the mixing entropy.

\section{Additional reorientation dynamics}

\begin{figure}
    \centering
    \includegraphics[width=0.85\linewidth]{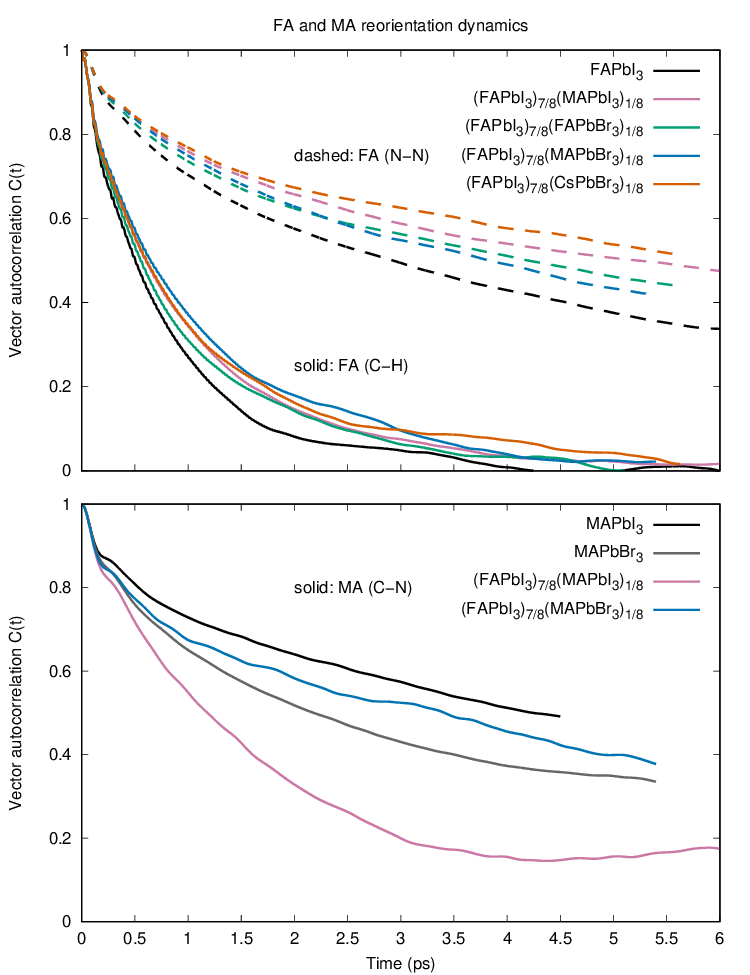}
    \caption{Vector autocorrelation functions $C(t)$ for A-site cation reorientation dynamics in the pure and mixed perovskites at 350~K. Top: FA-containing systems, comparing the FA N--N axis (dashed) and an additional FA internal C--H vector (solid). Bottom: MA-containing systems using the MA C--N axis. Autocorrelation functions were computed from
    AIMD trajectories and analyzed with TRAVIS$^{2}$.}
    \label{fig:rdyn_all}
\end{figure}

In the main text we report reorientation dynamics using the representative molecular axis for each A-site cation (FA: N--N; MA: C--N) to define species-resolved rotational correlation times. For completeness, Fig.~\ref{fig:rdyn_all} summarizes the corresponding vector autocorrelation functions \(C(t)\) for all studied compositions and for additional internal molecular
vectors. These curves illustrate the anisotropy of molecular motion within the inorganic cage and confirm that the qualitative trends in reorientational dynamics across compositions are robust with respect to the specific choice of molecular axis.

Because \(C(t)\) is normalized to \(C(0)=1\), we define the orientational correlation half-time \(t_{1/2}\) as the time at which \(C(t)\) first reaches 0.5, i.e., \(C(t_{1/2})=0.5\), and use \(t_{1/2}\) as a fit-free comparative measure of reorientation dynamics across systems. Using $t_{1/2}$ avoids imposing a specific functional form on $C(t)$, which can be ambiguous when multiple reorientation modes contribute or when the decay is non-exponential. All systems considered here reach $C(t)=0.5$ within the simulated time window, enabling a consistent comparison of reorientation dynamics across compositions (see Table~\ref{tab:correlation-half-time}). In the rotational-entropy estimator, $t_{1/2}$ is used as a proxy for the characteristic reorientation time $\tau$ because only time ratios enter the analysis.

\begin{table*}[th]
\setlength{\tabcolsep}{6pt}
\renewcommand{\arraystretch}{1.35}
\centering
\caption{Orientational correlation half-time \(t_{1/2}\) (ps) obtained from TRAVIS vector autocorrelation functions \(C(t)\). For FA, \(t_{1/2}\) corresponds to the N--N axis;
for MA, \(t_{1/2}\) corresponds to the C--N axis. Mixtures are reported for the
composition \(x=1/8\) used in the AIMD calculations.}
\begin{tabular}{
  @{} l                
  l                    
  S[table-format=2.5]  
  S[table-format=2.5]  
  @{}}
\toprule
\textbf{Mixing site} & \textbf{System} &
\multicolumn{1}{c}{\textbf{FA (N--N)}} &
\multicolumn{1}{c}{\textbf{MA (C--N)}} \\
& &
\multicolumn{1}{c}{\(t_{1/2}\) (ps)} &
\multicolumn{1}{c}{\(t_{1/2}\) (ps)} \\
\midrule

\multirow{3}{*}{Y site}
  & FAPbI$_3$      & 2.9250 & {} \\
  & FAPbBr$_3$     & 4.8340 & {} \\
  & \fapifapb{}    & 4.1950 & {} \\
\midrule

\multirow{3}{*}{A site}
  & FAPbI$_3$      & 2.9250 & {} \\
  & MAPbI$_3$      & {}      & 4.2450 \\
  & \fapimapi{}    & 5.2260 & 1.1830 \\
\midrule

\multirow{2}{*}{A and X site}
  & FAPbI$_3$      & 2.9250 & {} \\
  & CsPbBr$_3$     & {}      & {} \\
\addlinespace[2pt]
  & \fapicspb{}    & 6.0970 & {} \\
\midrule

\multirow{3}{*}{A and X site}
  & FAPbI$_3$      & 2.9250 & {} \\
  & MAPbBr$_3$     & {}      & 2.1820 \\
  & \fapimapb{}    & 3.8280 & 3.4130 \\
\bottomrule
\end{tabular}

\label{tab:correlation-half-time}
\end{table*}

\section{Estimation of thermodynamic functions error}
The mixing enthalpy per ABY$_3$ formula unit was computed from the
internal energies of the mixed and pure systems as
\begin{equation}
\Delta H_\text{mix}
=
\frac{\langle U_\text{mix}\rangle - x \langle U_A \rangle - (1-x) \langle U_B\rangle}{N_\text{fu}},
\label{eq:Hmix_def}
\end{equation}
where $U_\text{mix}$ is the total internal energy of the mixed
supercell, $U_A$ and $U_B$ are the internal energies of the pure
end members (e.g., FAPbI$_3$ and FAPbBr$_3$), $x$ is the mole fraction of the substituent/minority endmember $B$ in the mixture, and $N_\text{fu} = 64$ is the number of
ABY$_3$ formula units in the simulation cell. The internal energies $U_\text{mix}$, $U_A$, and $U_B$ were extracted as
time averages over the \textit{ab initio} molecular dynamics
trajectories. At fixed cell volume, we approximate $\Delta H_\text{mix}\approx \Delta U_\text{mix}$ per formula unit, as the $pV$ contribution is negligible on this scale.

Uncertainties were estimated from the energy time series by accounting for equilibration and time correlation, using an effective number of independent samples $N_\text{eff}$ to compute the standard error of the mean,
\(\mathrm{SEM}(U)=\sigma/\sqrt{N_\text{eff}}\), where $\sigma$ is the standard deviation of the (production) energy time series. The uncertainty in $\Delta H_\text{mix}$ was obtained by standard error propagation from the SEMs of the mixed and reference energies and is reported as one standard deviation.

Assuming uncorrelated fluctuations of the three energy estimators, the
uncertainty in the mixing energy
\begin{equation}
\Delta U_\text{mix} = U_\text{mix} - x U_A - (1-x) U_B
\end{equation}
is obtained by standard linear error propagation as
\begin{equation}
\mathrm{SEM}^2(\Delta U_\text{mix})
=
\mathrm{SEM}_\text{mix}^2
+ x^2 \mathrm{SEM}_A^2
+ (1-x)^2 \mathrm{SEM}_B^2,
\label{eq:variance_Emix}
\end{equation}
where $\mathrm{SEM}(\Delta U_\text{mix})$ is the standard error of the mean of $\Delta U_\text{mix}$.

The corresponding uncertainty in the mixing enthalpy per formula unit
follows directly from Eq.~\eqref{eq:Hmix_def}:
\begin{equation}
\mathrm{SEM}(\Delta H_\text{mix}) =
\frac{\mathrm{SEM}(\Delta U_\text{mix})}{N_\text{fu}}.
\label{eq:sigma_Hmix}
\end{equation}

The total Gibbs free energy of the mixed perovskites per ABY$_3$ formula unit is obtained as:
\begin{equation}
\Delta G_\text{mix}^\text{tot} = \Delta H_\text{mix} - T(\Delta S_\text{mix}^\text{conf} + \Delta S_\text{mix}^\text{rot}),
\label{eq:Gmix_def}
\end{equation}
where $\Delta H_\text{mix}$ is the mixing enthalpy per formula unit,
$T$ is the simulation temperature, and $\Delta S_\text{mix}^\text{conf}$ and $\Delta S_\text{mix}^\text{rot}$ are the
conformational and rotational entropies of mixing per formula unit.

Since $\Delta S_\text{mix}^\text{conf}$ is evaluated analytically, it carries no statistical uncertainty.

The rotational entropy of mixing is estimated from orientational correlation functions via the fit-free half-time $t_{1/2}$ (Table~\ref{tab:correlation-half-time}), which we use as an operational proxy for the characteristic reorientation time scale. Because $t_{1/2}$ is obtained from the first crossing $C(t_{1/2})=0.5$, it avoids imposing a specific functional form on $C(t)$. However, a rigorous statistical uncertainty for $\Delta S_\text{mix}^\text{rot}$ is still nontrivial to assign for the present AIMD trajectory lengths. In particular, $t_{1/2}$ depends on the discrete sampling of $C(t)$, and exhibits block-to-block fluctuations for $\sim$10--20~ps trajectories.

Importantly, the rotational-entropy term is numerically small compared with the dominant configurational-entropy contribution and does not affect the qualitative thermodynamic conclusions. Consequently, even conservative variations in $\Delta S_\text{mix}^\text{rot}$ would not change the sign of $\Delta G_\text{mix}^\text{tot}$ nor the absence of a miscibility gap inferred from the curvature analysis. We therefore report $\Delta S_\text{mix}^\text{rot}$ as a best estimate and do not propagate a formal uncertainty for this term.

Since $\Delta S_\text{mix}^\text{conf}$ is analytical and we do not propagate uncertainties for $\Delta S_\text{mix}^\text{rot}$, the reported uncertainty of the total mixing free energy reduces to
\begin{equation}
\mathrm{SEM}(\Delta G_\text{mix}^\text{tot}) \approx \mathrm{SEM}(\Delta H_\text{mix}) .
\end{equation}

\section{Regular solution analysis}

\begin{figure}
    \centering
    \includegraphics[width=0.8\linewidth]{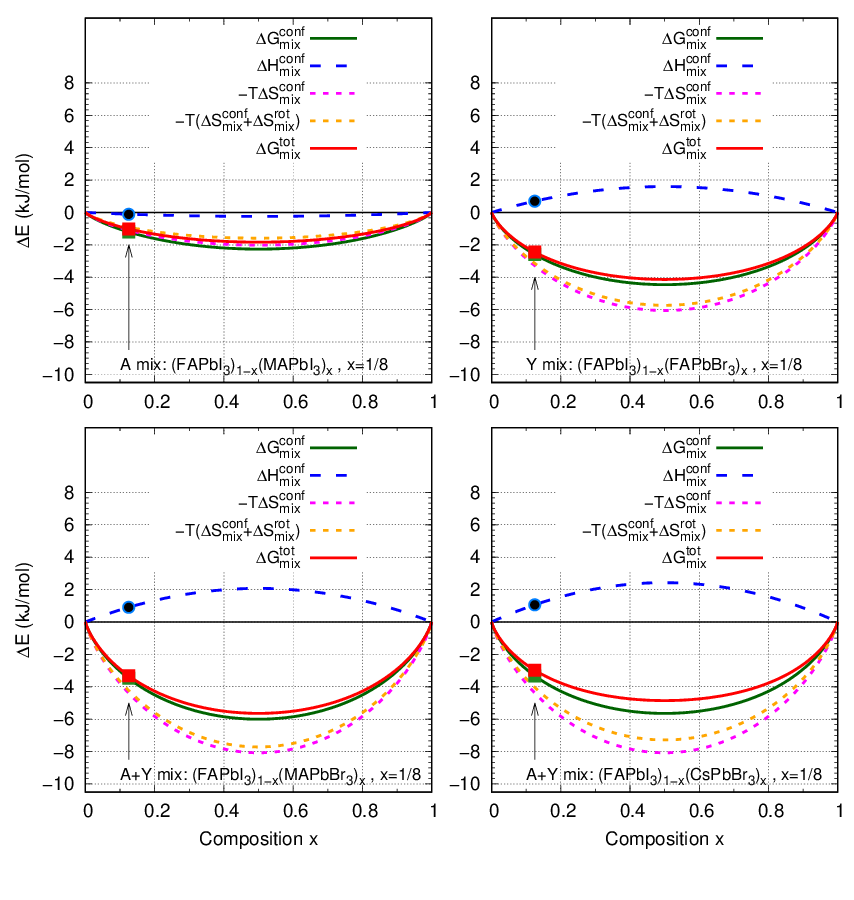}
\caption{Regular-solution model analysis of the mixing free energy at 350~K.
Shown are $\Delta H_\text{mix}(x)=W x(1-x)$, the configurational entropy term
$-T\Delta S_\text{mix}^\text{conf}(x)$ with $n_s$ mixed sites per formula unit,
the combined entropic term $-T(\Delta S_\text{mix}^\text{conf}+\Delta S_\text{mix}^\text{rot})$
using $-T\Delta S_\text{mix}^\text{rot}(x)=T\lambda x(1-x)$, and the corresponding
free energies $\Delta G_\text{mix}^\text{conf}(x)$ and $\Delta G_\text{mix}^\text{tot}(x)$.
Parameters $W$ and $\lambda$ are obtained from the AIMD-evaluated thermodynamic
quantities at $x=1/8$.}
\label{regularsolutionmodel}
\end{figure}

To assess the stability of the solid solutions against phase separation, we analyzed the curvature of the mixing free energy within a regular-solution model, parameterized using the AIMD-evaluated thermodynamic quantities at $x=1/8$ (Fig.~\ref{regularsolutionmodel}).

In the regular-solution approximation, the enthalpy of mixing per ABY$_3$ formula unit is represented by a composition-independent bowing parameter $W$,
\begin{equation}
\Delta H_\text{mix}(x)=W\,x(1-x),
\end{equation}
with
\begin{equation}
W=\frac{\Delta H_\text{mix}(x)}{x(1-x)}.
\end{equation}
The configurational entropy of mixing is treated in the ideal-solution approximation on the relevant mixed sublattice(s),
\begin{equation}
\Delta S_\text{mix}^\text{conf}(x) = -n_s k_B \left[x\ln x+(1-x)\ln(1-x)\right],
\end{equation}
where $n_s$ is the number of mixed sites per ABY$_3$ formula unit ($n_s=1$
for A-mixing, $n_s=3$ for Y-mixing, and $n_s=4$ for simultaneous A+Y mixing).

For systems containing molecular A-site cations, we include an additional rotational-entropy correction using the same symmetric regular-solution form as in the main text,
\begin{equation}
-T\Delta S_\text{mix}^\text{rot}(x) = T\lambda\,x(1-x),
\label{eq:rot_regular_solution}
\end{equation}
where $\lambda$ is determined from the AIMD-evaluated rotational entropy of mixing at $x$:
\begin{equation}
\lambda = \frac{-\Delta S_\text{mix}^\text{rot}(x)}{x(1-x)}.
\end{equation}
With this definition, $\lambda>0$ corresponds to a loss of rotational entropy
upon mixing (i.e., $\Delta S_\text{mix}^\text{rot}<0$).

Combining the enthalpic and rotational contributions leads to an effective
regular-solution parameter,
\begin{equation}
W_\mathrm{eff}=W+T\lambda,
\end{equation}
and the corresponding total mixing free energy becomes
\begin{equation}
\Delta G_\text{mix}^\text{tot}(x)
=
W_\mathrm{eff}x(1-x)
+ n_s k_B T\left[x\ln x+(1-x)\ln(1-x)\right].
\label{eq:regular_solution_total}
\end{equation}
For comparison, the configurational-only free energy is recovered by setting
$\lambda=0$ (i.e., $W_\mathrm{eff}=W$), yielding $\Delta G_\text{mix}^\text{conf}(x)$.

Within this model, a miscibility gap occurs when the curvature of
$\Delta G_\text{mix}(x)$ becomes negative, which leads to the condition
\begin{equation}
W_\mathrm{eff} > 2 n_s k_B T .
\end{equation}
The resulting composition-dependent profiles of $\Delta H_\text{mix}$,
$-T\Delta S_\text{mix}^\text{conf}$, $-T(\Delta S_\text{mix}^\text{conf}
+\Delta S_\text{mix}^\text{rot})$, and the corresponding
$\Delta G_\text{mix}^\text{conf}$ and $\Delta G_\text{mix}^\text{tot}$ are
shown in Fig.~\ref{regularsolutionmodel}.

\section{Hydrogen bond analysis}

\begin{figure*}[!p]
\centering
\subfloat[\textbf{\fapimapb{}}]{\label{fig_cdf_1_FAPI-MAPB}
\begin{tabular}[t]{@{}c@{}}
\includegraphics[width=.235\linewidth]{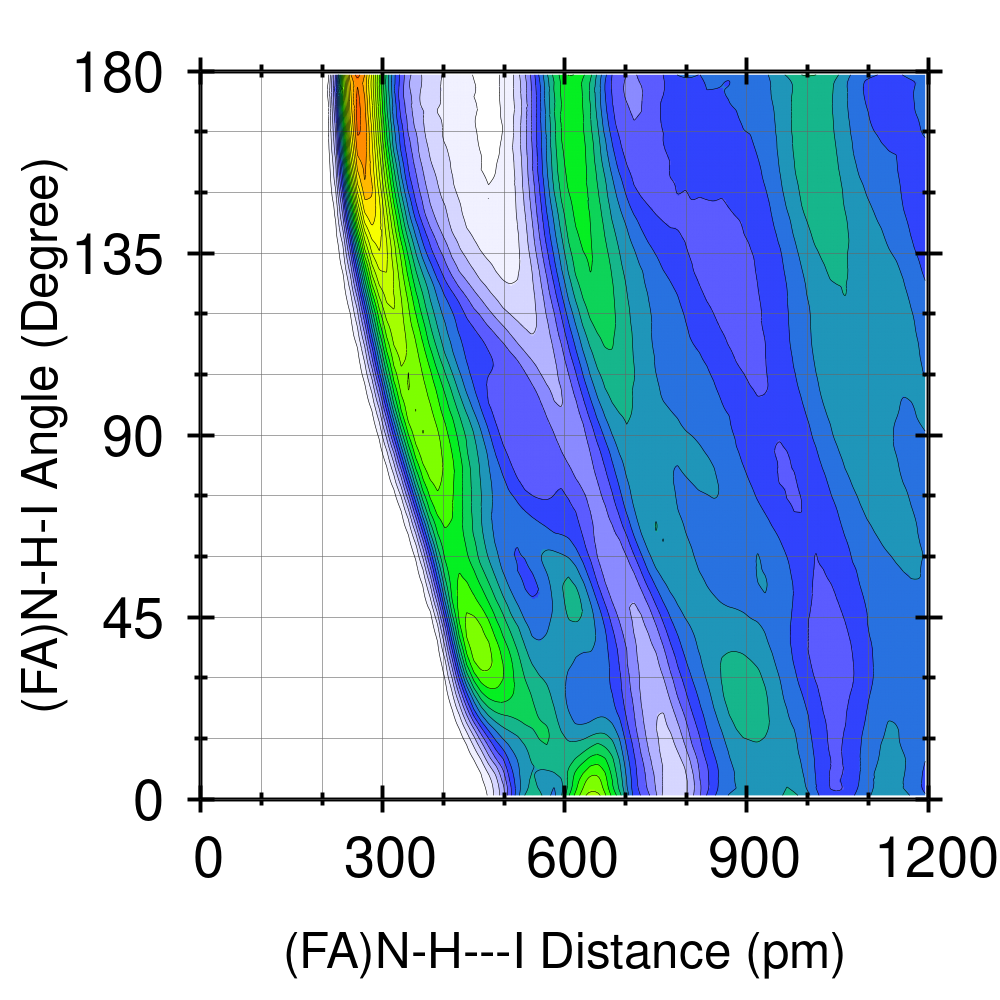}%
\includegraphics[width=.235\linewidth]{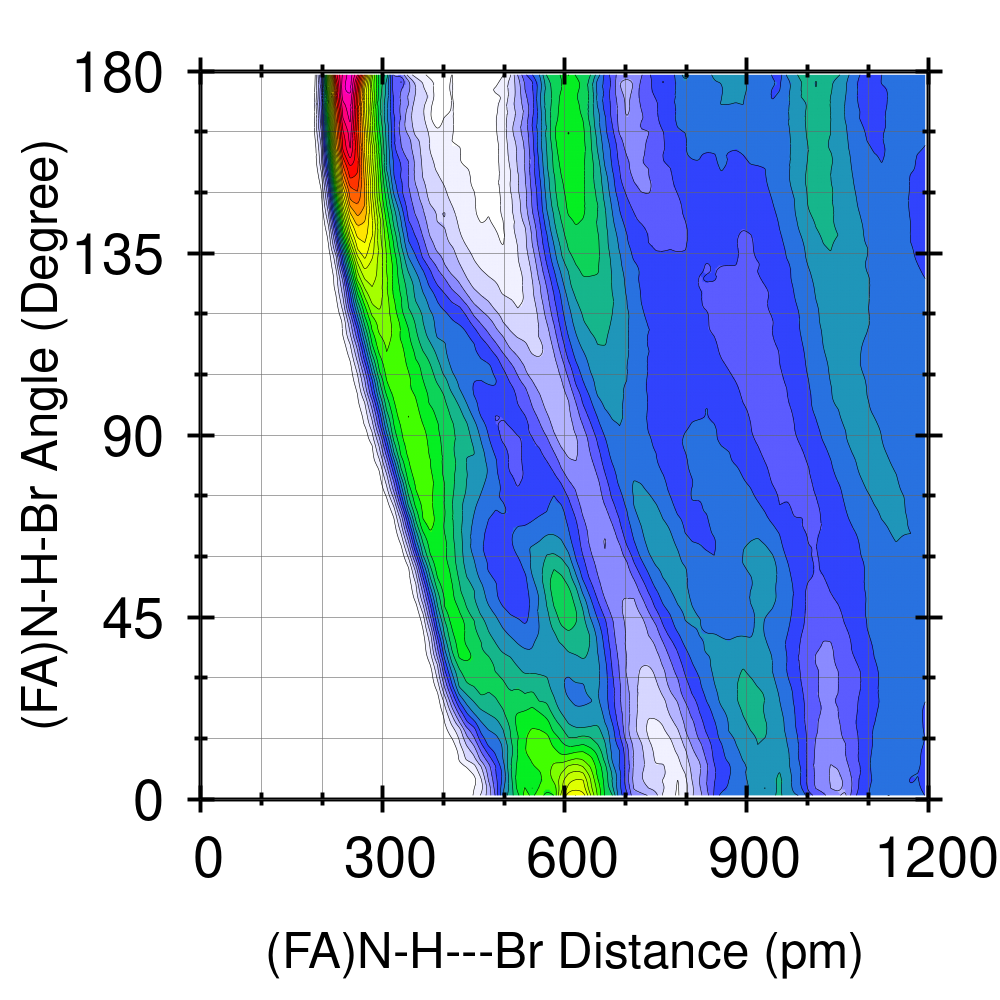}
\includegraphics[width=.235\linewidth]{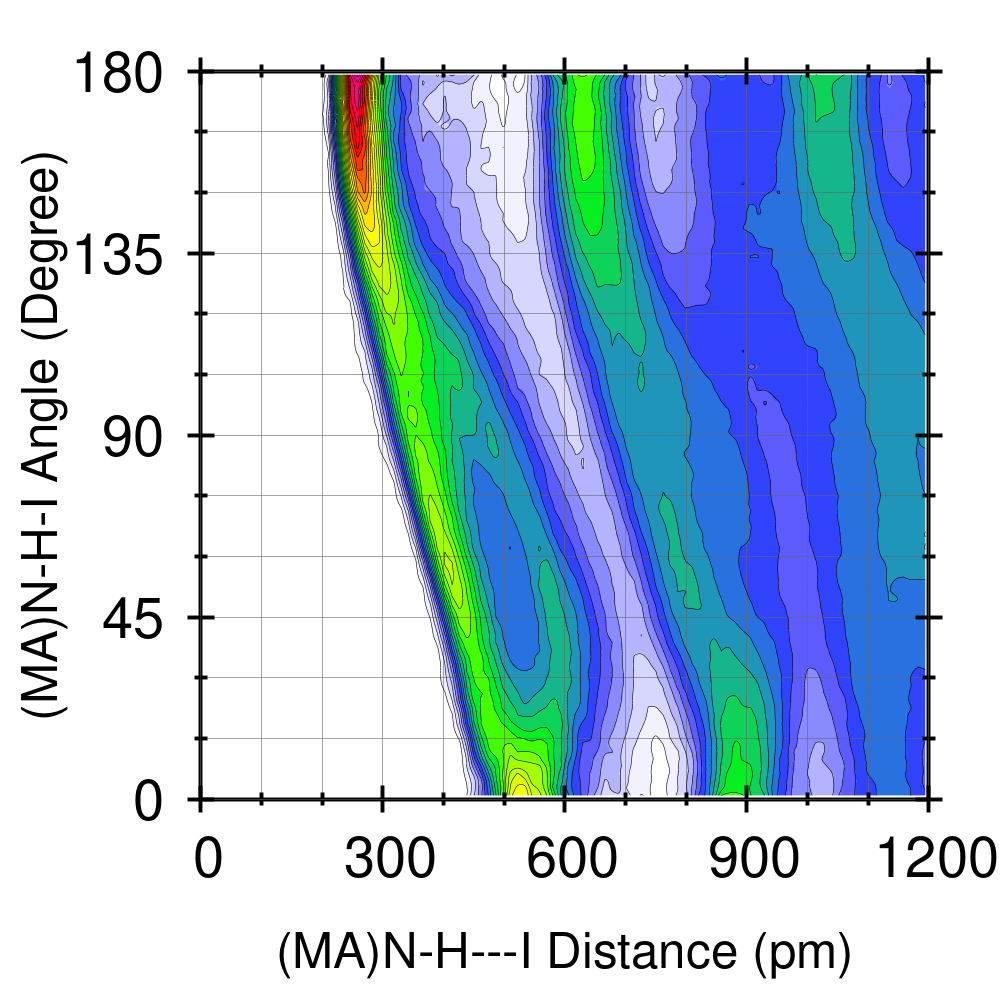}%
\includegraphics[width=.235\linewidth]{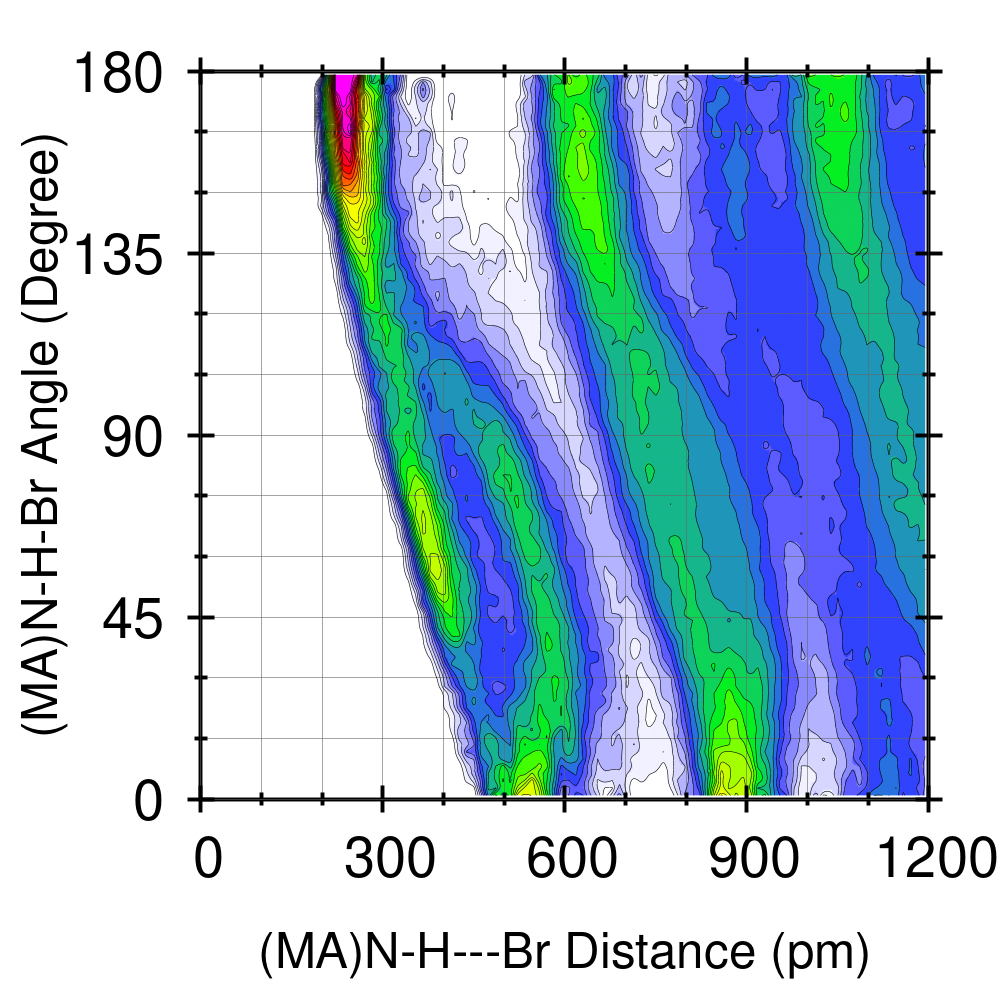}
\end{tabular}%
}
\\
\subfloat[\textbf{\fapifapb{}}]{\label{fig_cdf_5_FAPI-FAPB}
\begin{tabular}[t]{@{}c@{}}
\includegraphics[width=.235\linewidth]{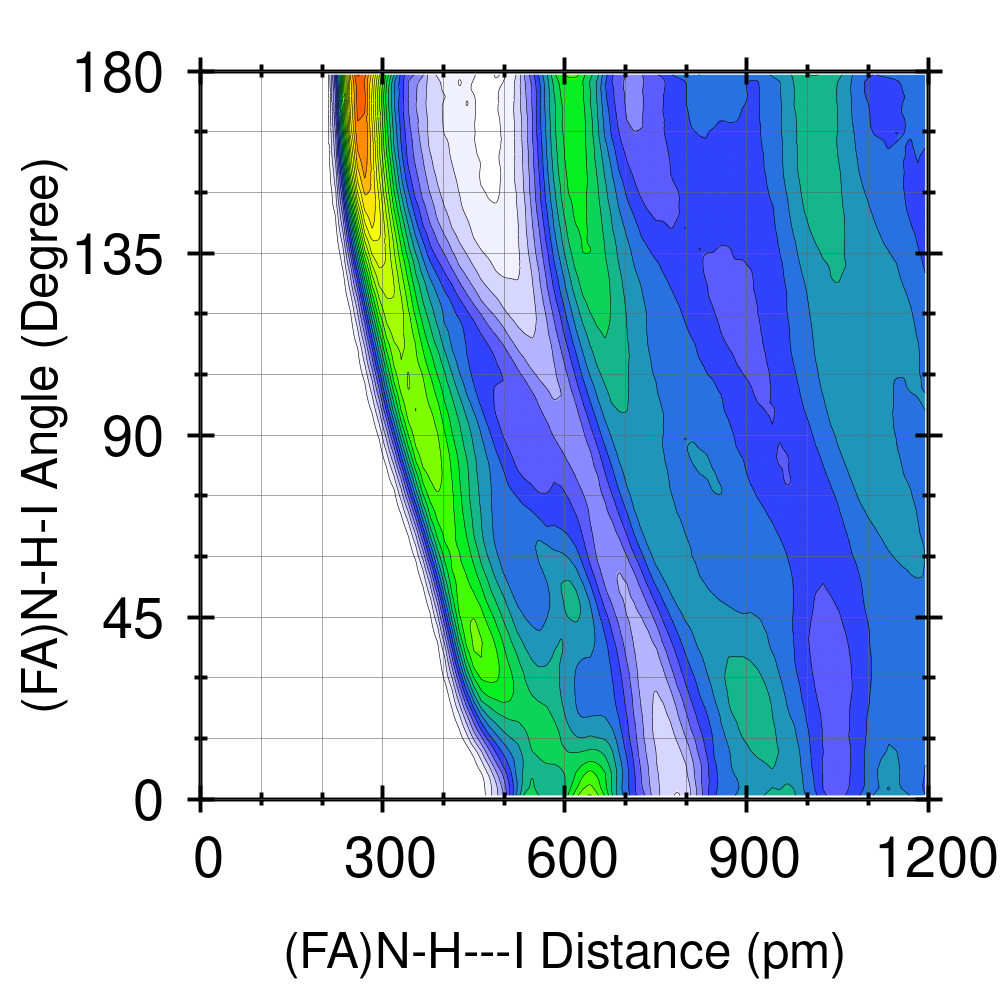}%
\includegraphics[width=.235\linewidth]{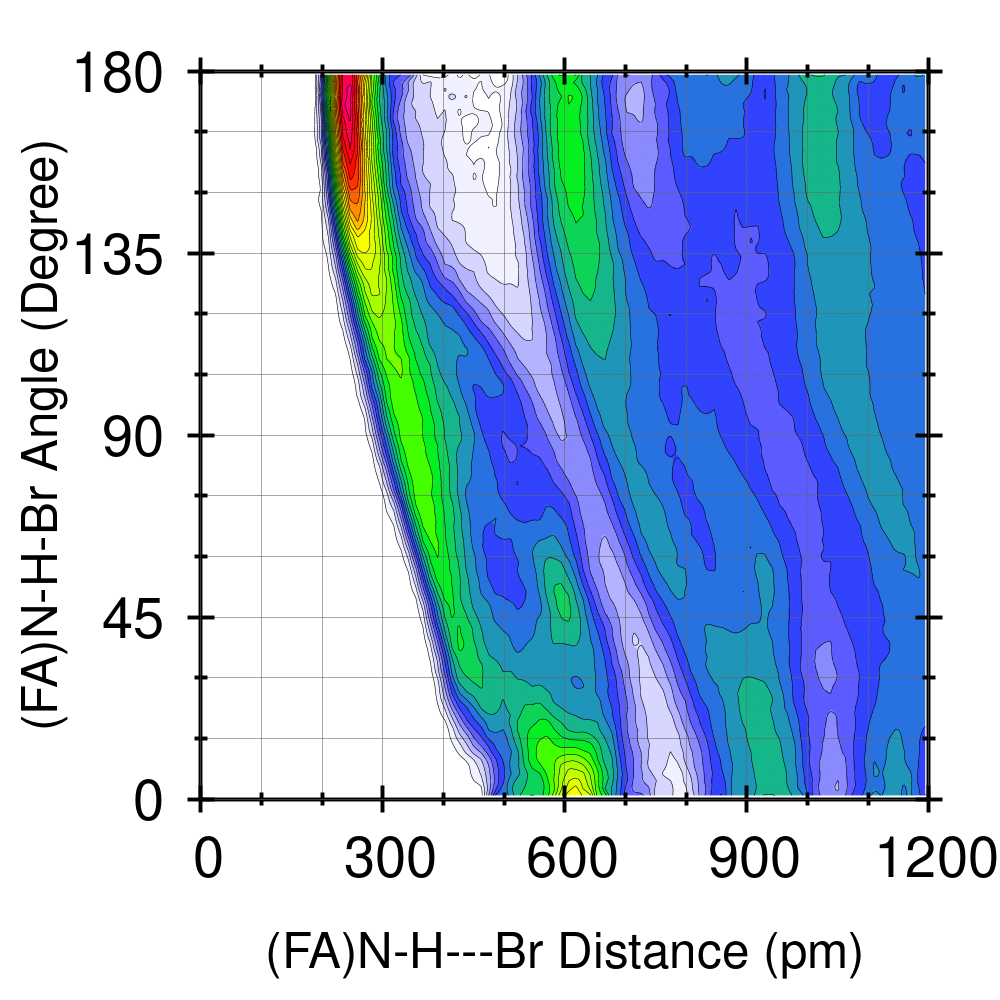}\\[-3pt]
\end{tabular}
}
\subfloat[\textbf{\fapimapi{}}]{\label{fig_cdf_6_FAPI-MAPI}
\begin{tabular}[b]{@{}c@{}}
\includegraphics[width=.235\linewidth]{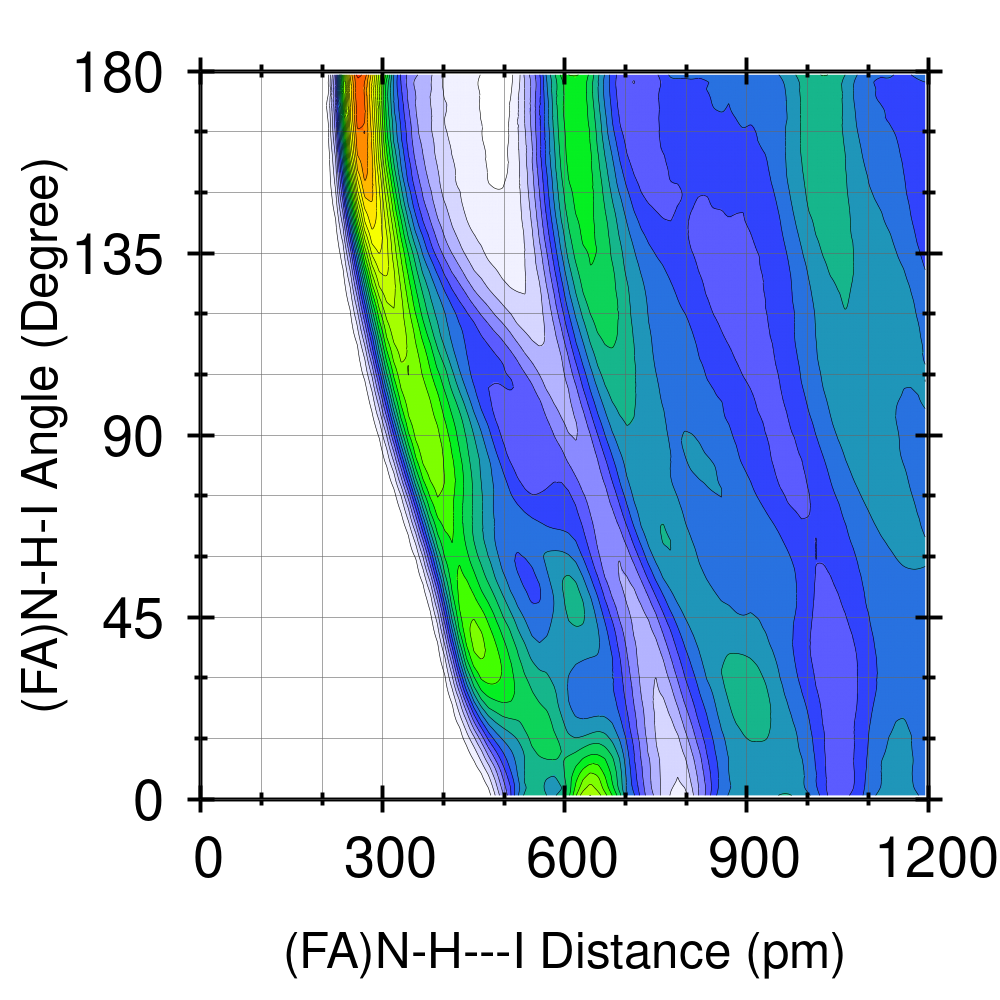}%
\includegraphics[width=.235\linewidth]{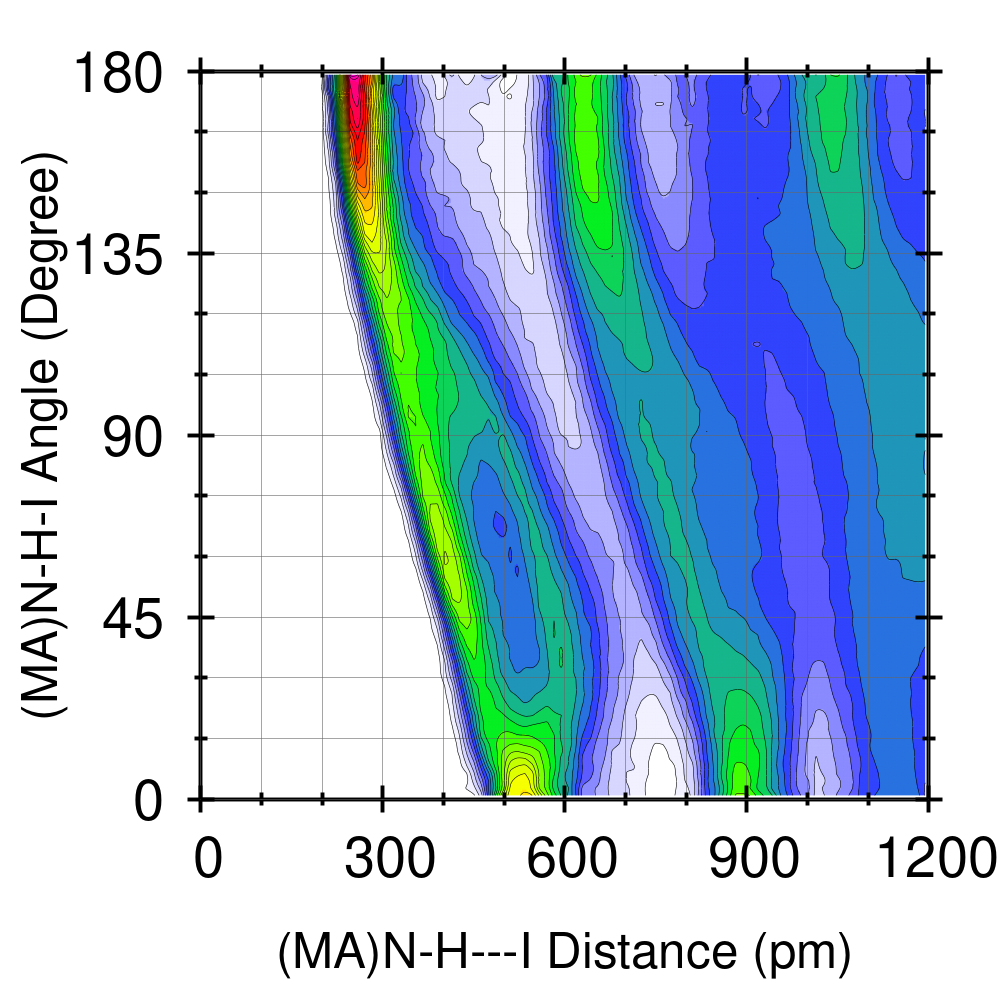}
\end{tabular}
}
\\
\subfloat[\textbf{\fapicspb{}}]{\label{fig_cdf_7_FAPI-CsPB}
\begin{tabular}[b]{@{}c@{}}
\includegraphics[width=.235\linewidth]{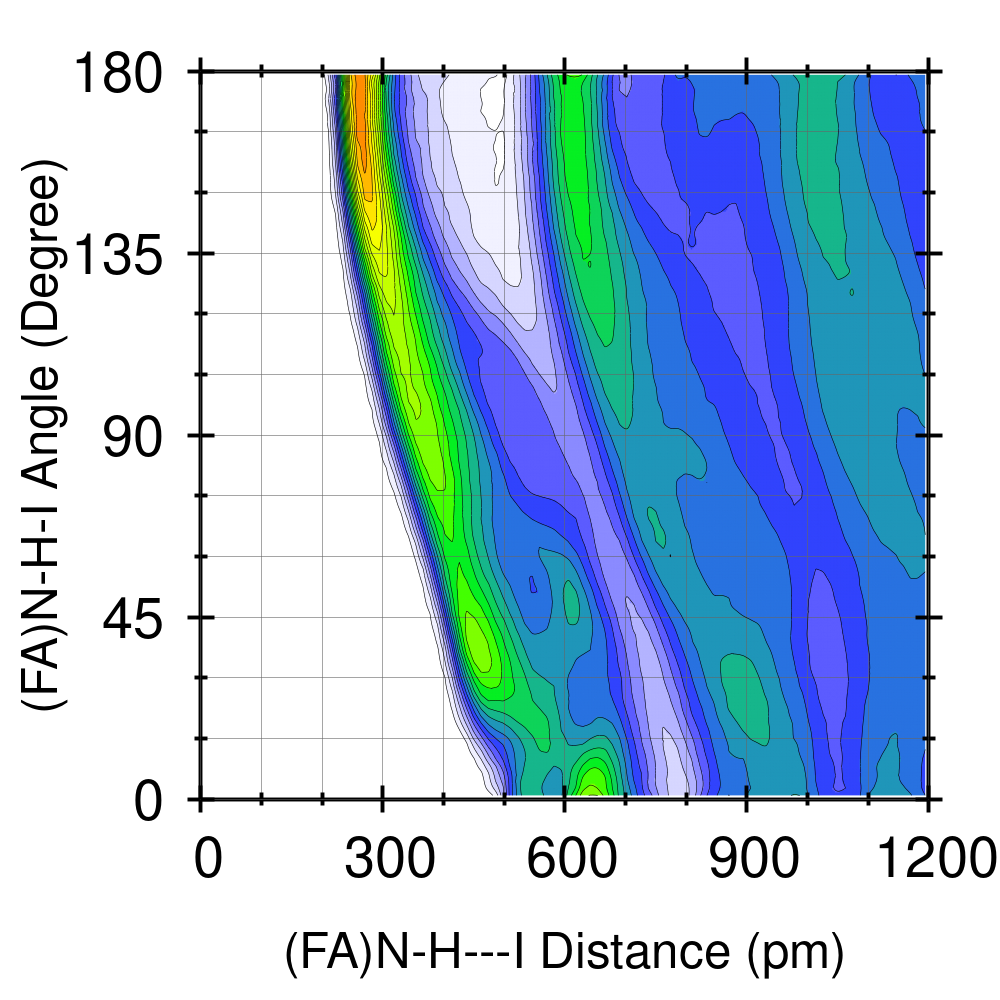}%
\includegraphics[width=.235\linewidth]{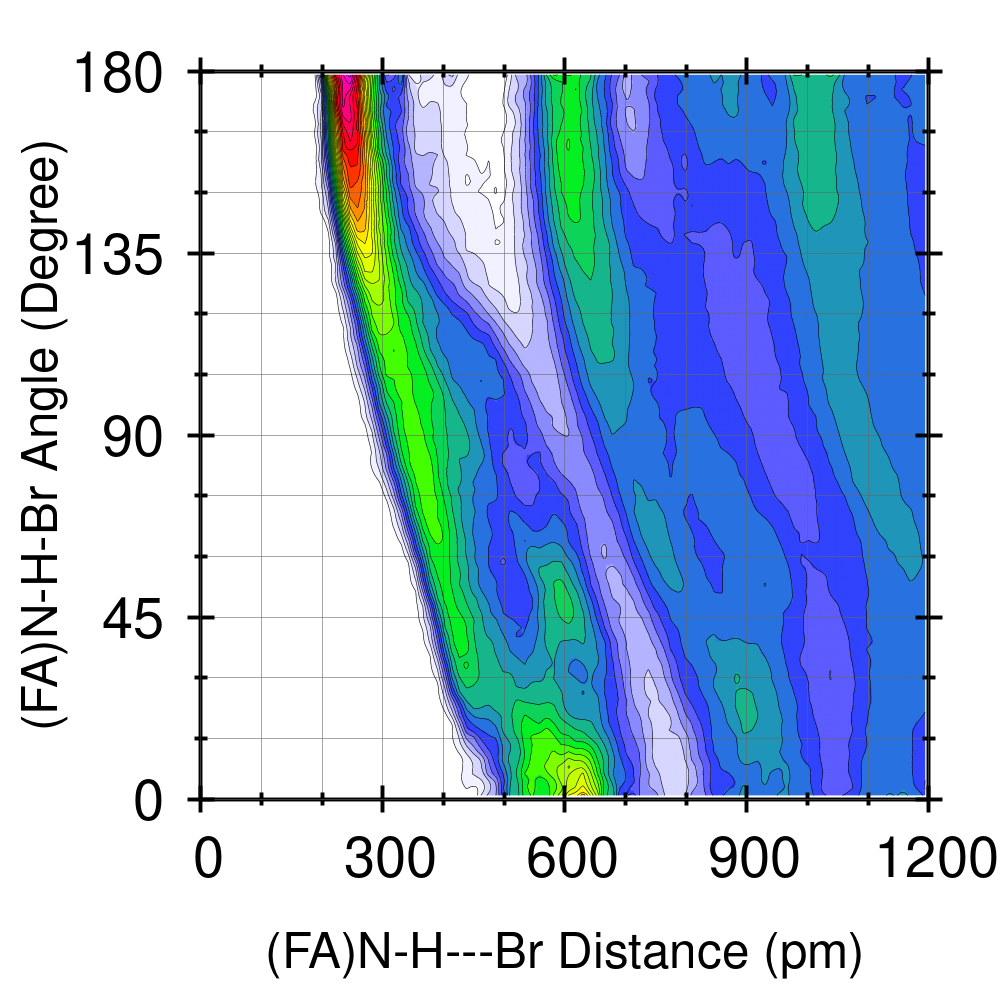}
\end{tabular}
}
\begin{tabular}[t]{@{}c@{}}
\includegraphics[width=.4\linewidth]{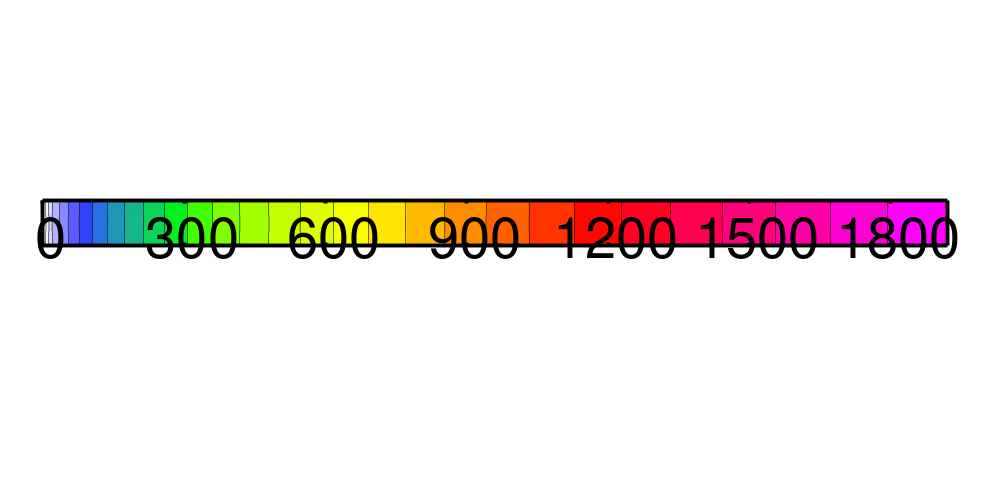}%
\end{tabular}
\caption{Combined distribution functions (CDFs) for \hb{N}{I} and \hb{N}{Br} hydrogen bonds (HBs) in the studied perovskites. The CDFs are shown as 2D histograms where the color scale indicates the configuration frequency in each of the \(300\times300\) bins (proportional to the configuration probability). Panels (a--d) correspond to the mixed compounds: (a) \fapimapb{}, (b) \fapifapb{}, (c) \fapimapi{}, and (d) \fapicspb{}. In each mixed system, the \hb{N}{I} and \hb{N}{Br} CDFs are reported separately for FA and MA whenever present; for \fapicspb{} only FA contributes.}
\label{fig_cdf_total_v3-mix}
\end{figure*}

\begin{figure*}[!p]
\raggedright
\subfloat[\textbf{\fapi{}}]{\label{fig_cdf_2_FAPI}\includegraphics[width=.235\linewidth]{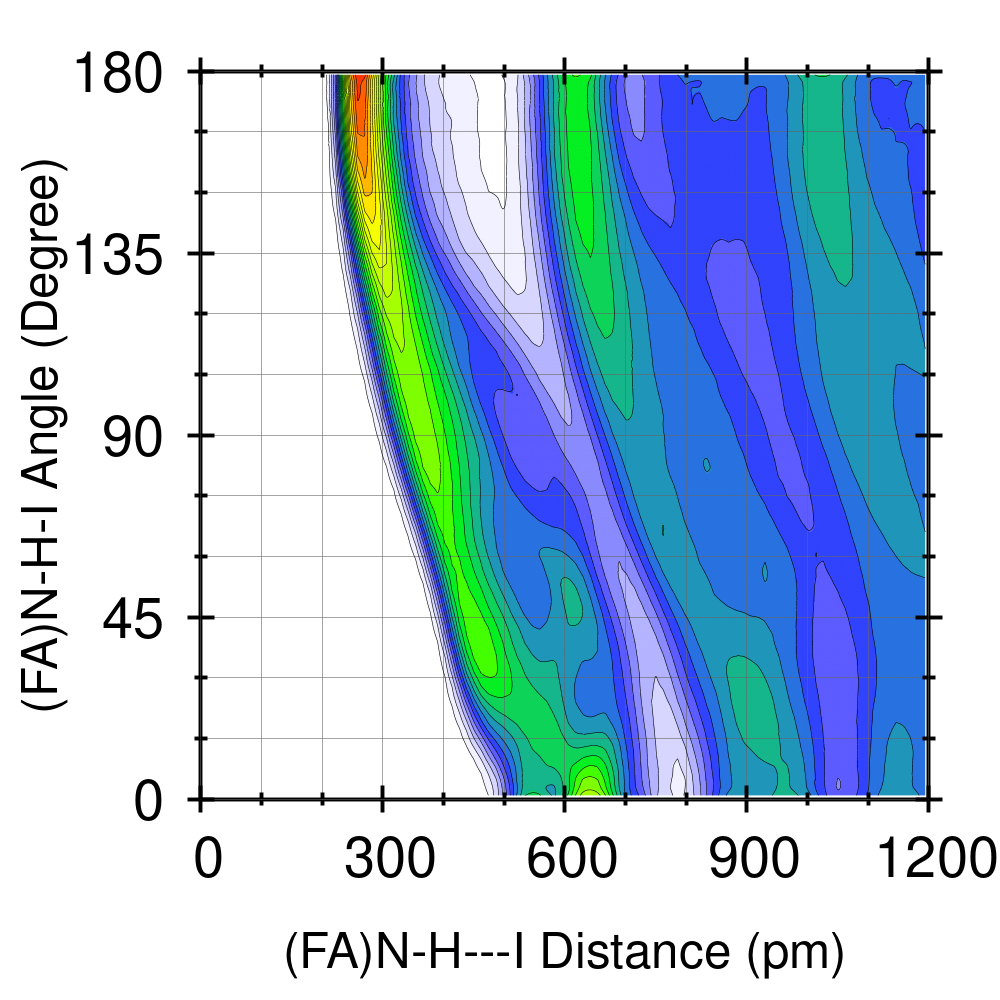}}
\subfloat[\textbf{\fapb{}}]{\label{fig_cdf_4_FAPB}\includegraphics[width=.235\linewidth]{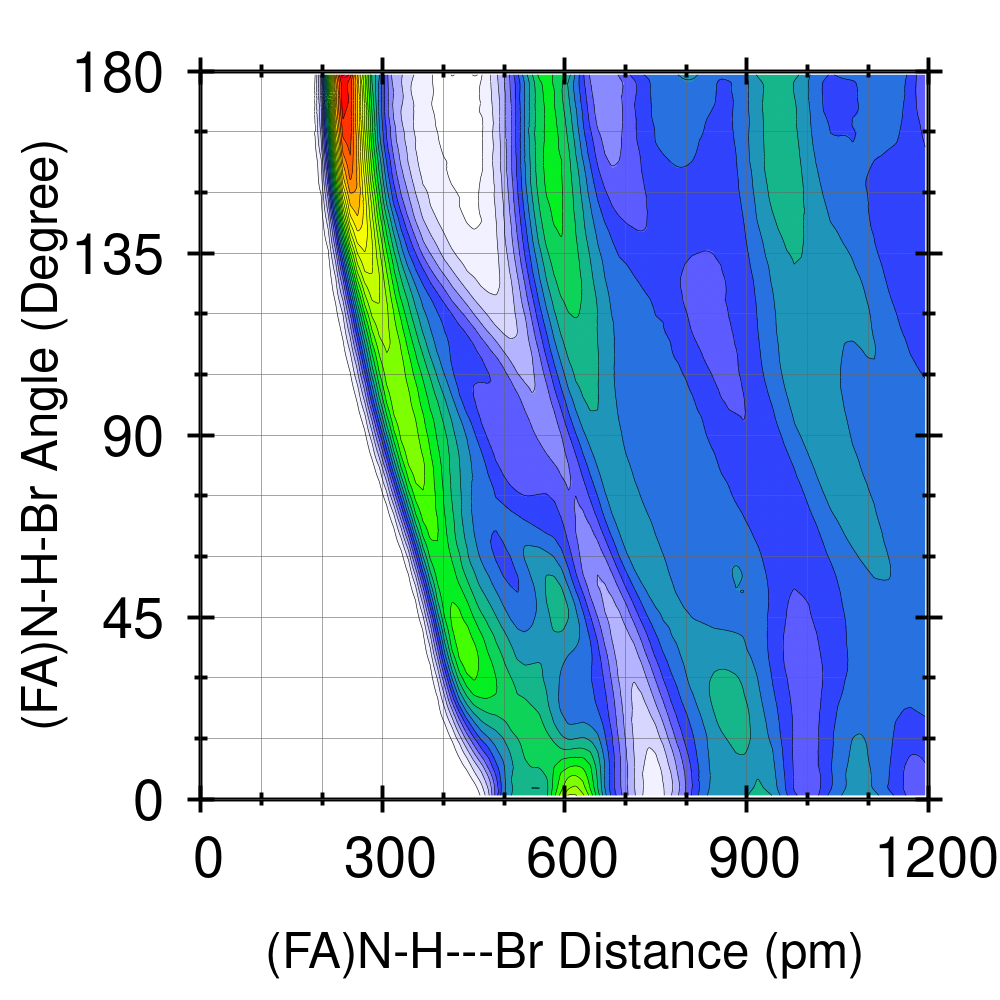}} 
\\
\subfloat[\textbf{\mapi{}}]{\label{fig_cdf_8_MAPI}\includegraphics[width=.235\linewidth]{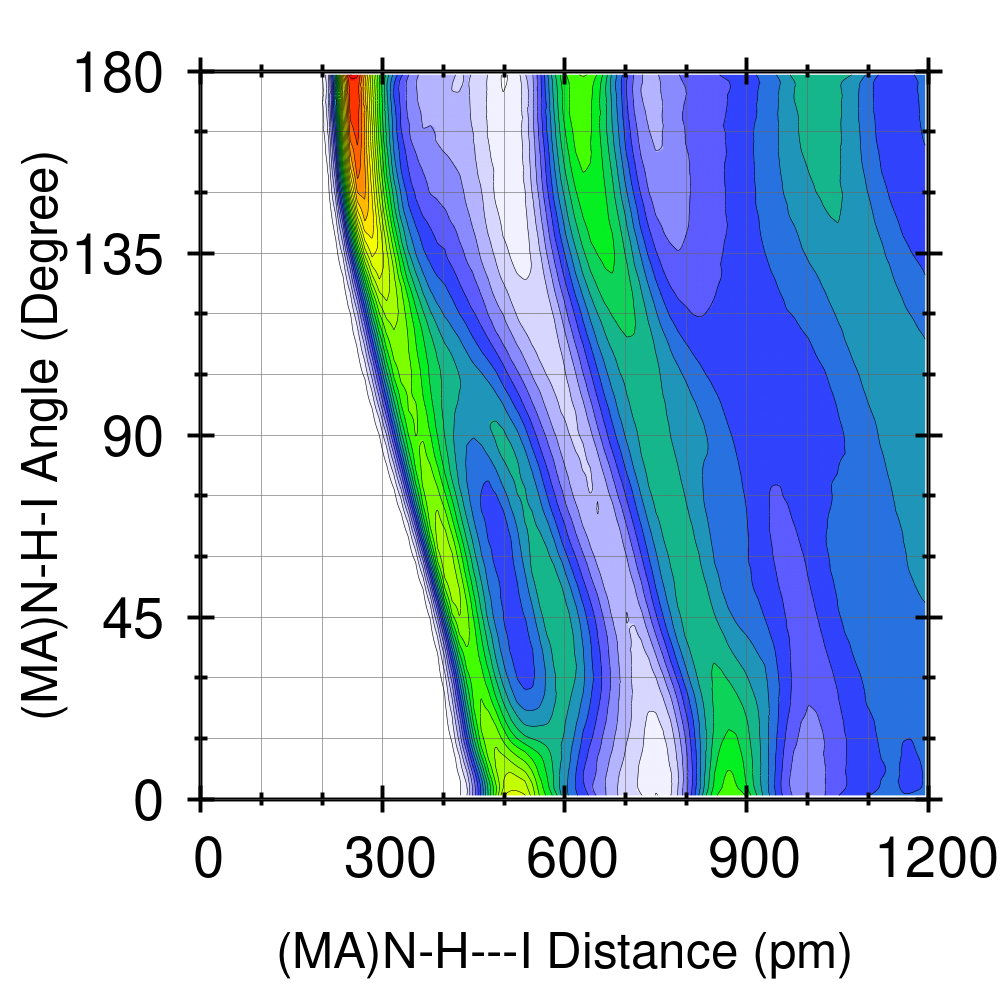}}
\subfloat[\textbf{\mapb{}}]{\label{fig_cdf_3_MAPB}\includegraphics[width=.235\linewidth]{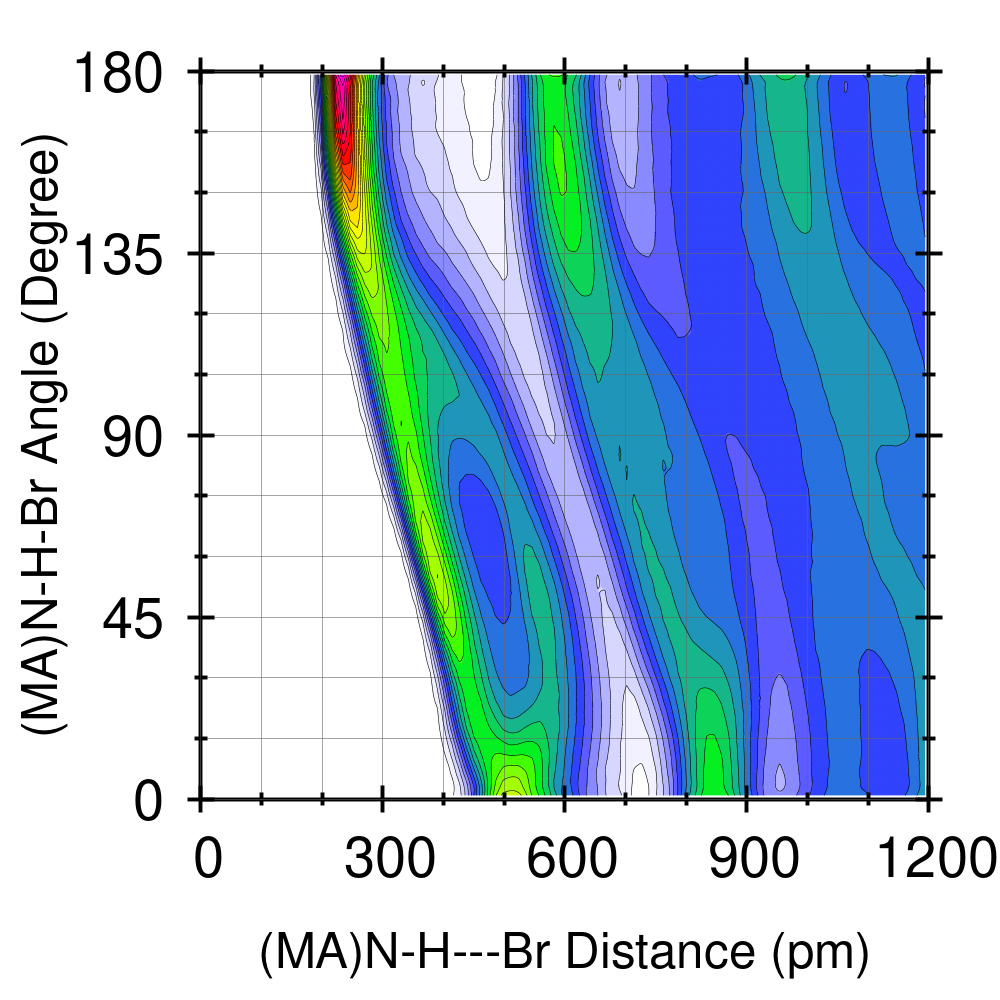}} \qquad\quad
\subfloat[\textbf{z axi box}]{\label{fig_cdf_box_1_FAPI-MAPB}
\begin{tabular}[t]{@{}c@{}}
\includegraphics[width=.4\linewidth]{Pictures/1_FAPI-MAPB/cdf/cdf_FAPI-MAPB_I-FA_box_v3.png}%
\end{tabular}%
}
\caption{Combined distribution functions (CDFs) for \hb{N}{I} and \hb{N}{Br} hydrogen bonds (HBs) in the studied perovskites. The CDFs are shown as 2D histograms where the color scale indicates the configuration frequency in each of the \(300\times300\) bins (proportional to the configuration probability). Panels (a--d) show the corresponding pure end members: (e) \fapi{}, (f) \fapb{}, (g) \mapi{}, and (h) \mapb{}.}
\label{fig_cdf_total_v3-pure}
\end{figure*}

Figures~\ref{fig_cdf_total_v3-mix}--\ref{fig_cdf_total_v3-pure} report the CDF maps of $d(\mathrm{H}-Y)$ versus $\measuredangle(\mathrm{N}-\mathrm{H}-Y)$ (with $Y=\mathrm{I},\mathrm{Br}$) for the mixed and pure perovskites. In all cases, the geometrical hydrogen-bond (HB) definition used in the main text ($d(\mathrm{H}-Y)\le 3$~\AA{} and $135^{\circ}\le \measuredangle(\mathrm{N}-\mathrm{H}-Y)\le 180^{\circ}$) encloses the high-probability region of the CDFs, supporting its transferability across compositions for extracting HB dynamical properties.

\begin{figure*}[!tbp]
  \centering
  \begin{minipage}[b]{0.49\textwidth}
    \includegraphics[width=\textwidth]{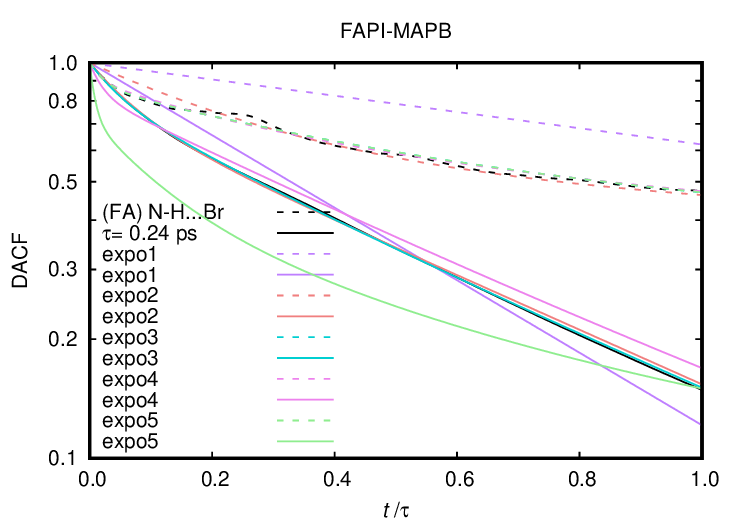}
  \end{minipage}
  \begin{minipage}[b]{0.49\textwidth}
    \includegraphics[width=\textwidth]{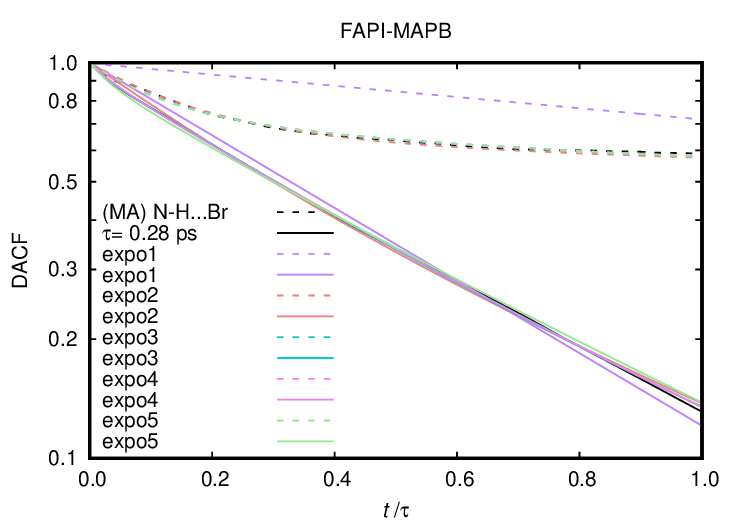}
  \end{minipage}
  \begin{minipage}[b]{0.49\textwidth}
    \includegraphics[width=\textwidth]{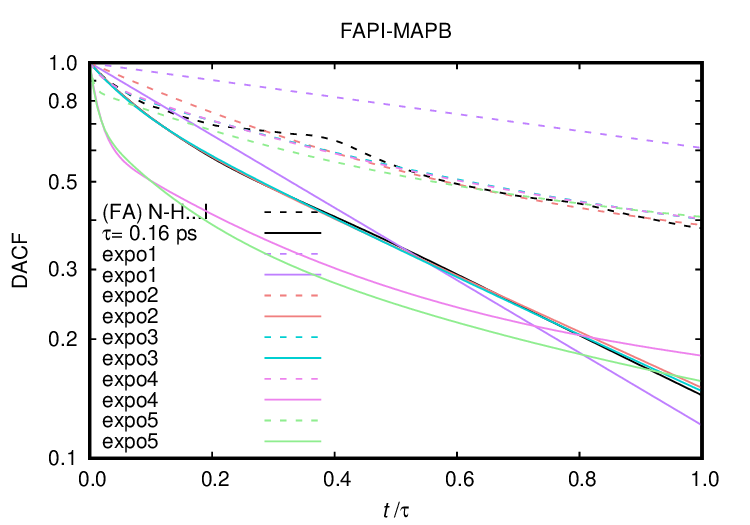}
  \end{minipage}
  \begin{minipage}[b]{0.49\textwidth}
    \includegraphics[width=\textwidth]{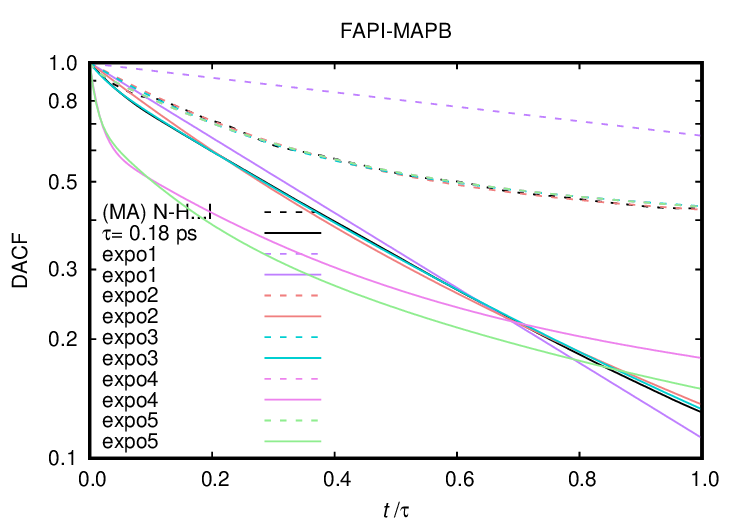}
  \end{minipage}
  \begin{minipage}[b]{0.49\textwidth}
    \includegraphics[width=\textwidth]{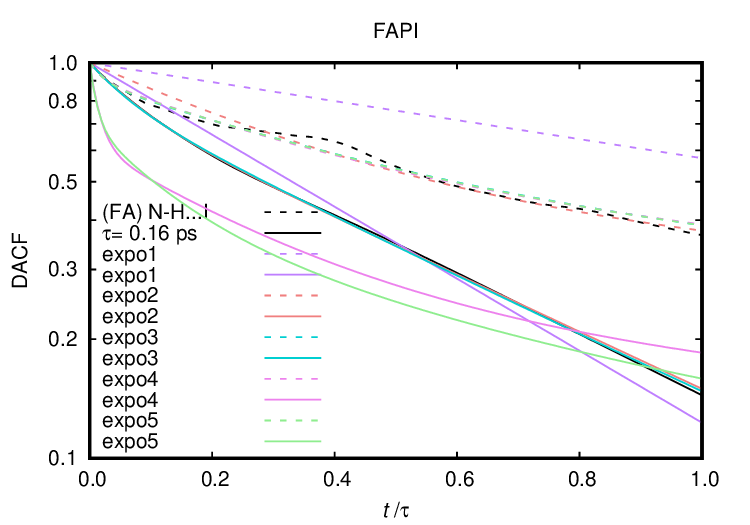}
  \end{minipage}
  \begin{minipage}[b]{0.49\textwidth}
    \includegraphics[width=\textwidth]{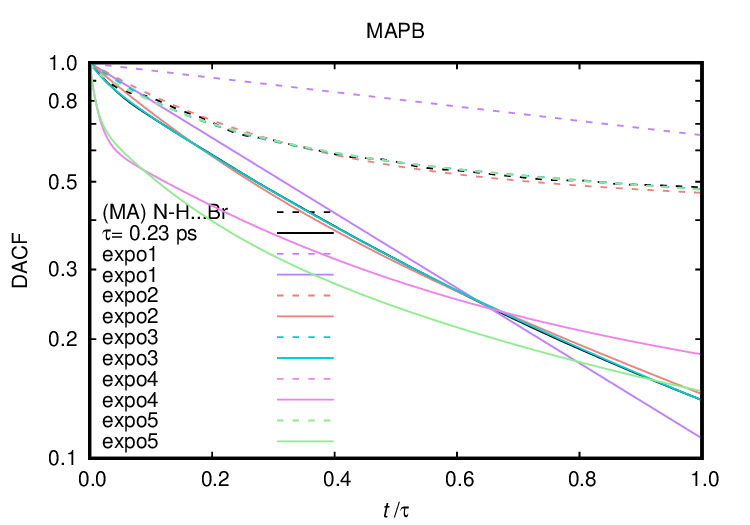}
  \end{minipage}
\caption{Hydrogen-bond autocorrelation functions (ACFs) for \hb{N}{I} and \hb{N}{Br} HBs in \fapimapb{}, \fapi{}, and \mapb{}. Solid lines (black): continuous ACF; dotted lines (black): intermittent ACF. Colored curves show poly-exponential fits using 1--5 exponential terms (as indicated in the legend) applied to the continuous and intermittent ACF decay. These ACFs underpin the HB lifetime analysis discussed in the main text.}
\label{fig_acf_FAPI-MAPB}
\end{figure*}

\begin{figure*}[!tbp]
  \centering 
  \begin{minipage}[b]{0.49\textwidth}
    \includegraphics[width=\textwidth]{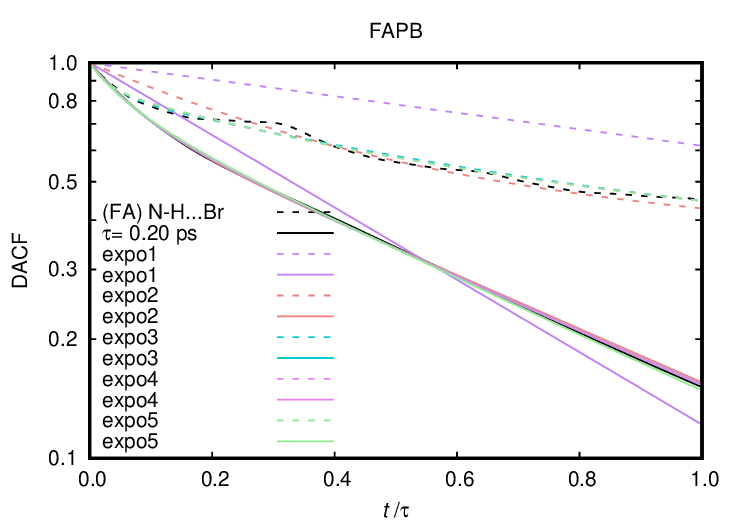}
  \end{minipage}
  \begin{minipage}[b]{0.49\textwidth}
    \includegraphics[width=\textwidth]{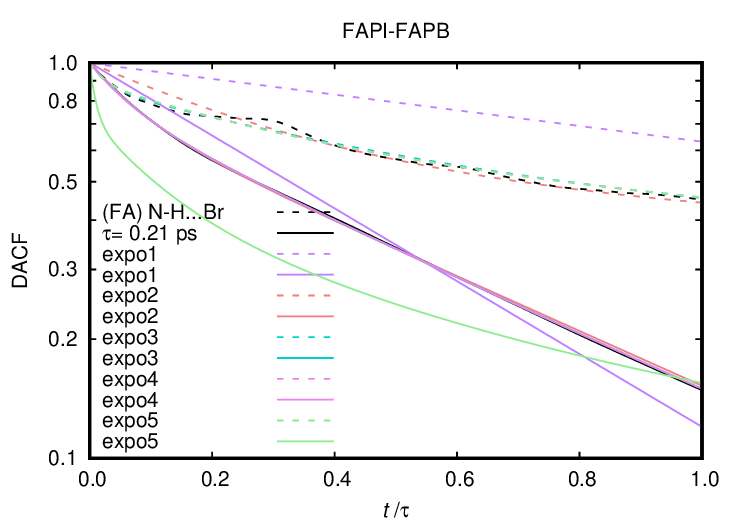}
  \end{minipage}
  \begin{minipage}[b]{0.49\textwidth}
    \includegraphics[width=\textwidth]{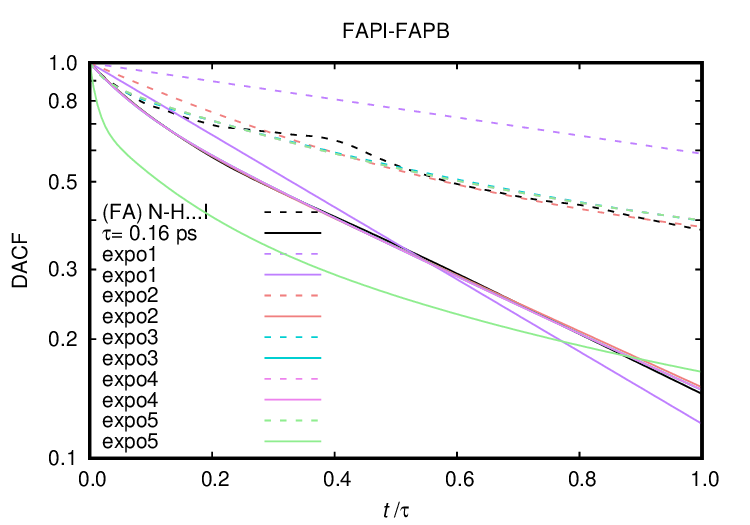}
  \end{minipage}
  \begin{minipage}[b]{0.49\textwidth}
    \includegraphics[width=\textwidth]{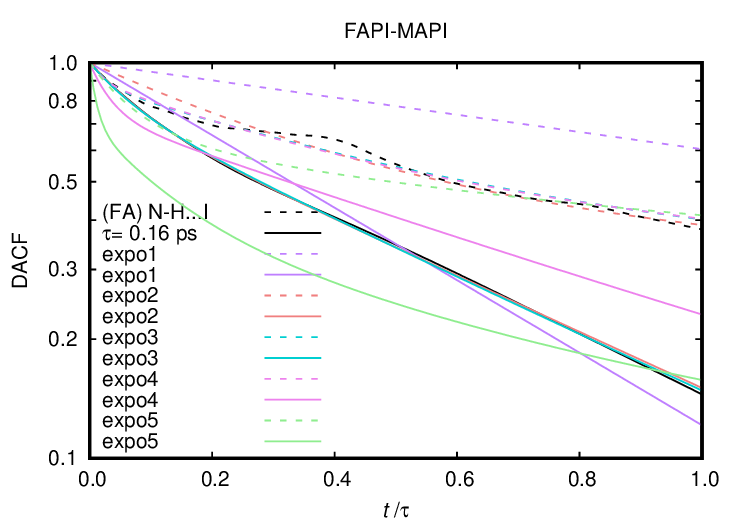}
  \end{minipage}
  \begin{minipage}[b]{0.49\textwidth}
    \includegraphics[width=\textwidth]{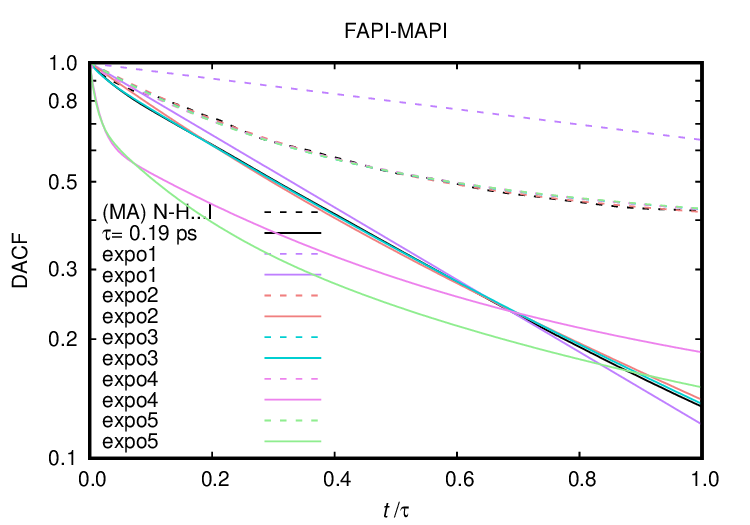}
  \end{minipage}
  \begin{minipage}[b]{0.49\textwidth}
    \includegraphics[width=\textwidth]{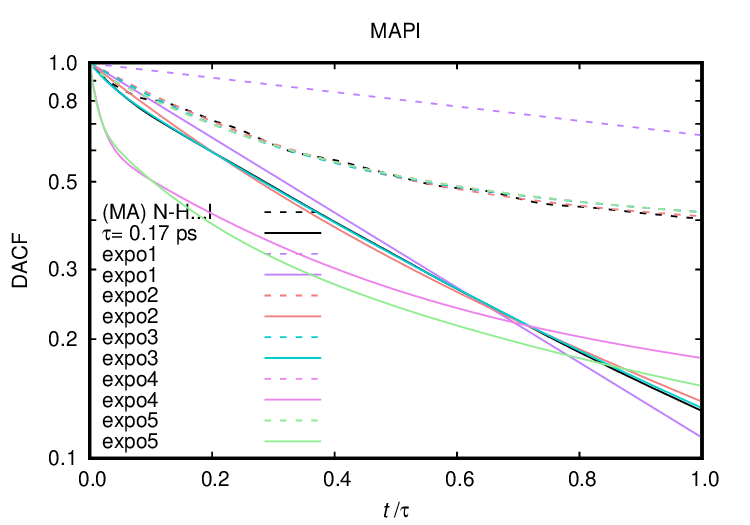}
  \end{minipage}
\caption{Hydrogen-bond autocorrelation functions (ACFs) for \hb{N}{I} and/or \hb{N}{Br} HBs in \fapb{}, \fapifapb{}, \fapimapi{}, \mapi{}, \fapb{} . Solid lines (black): continuous ACF; dotted lines (black): intermittent ACF. Poly-exponential fits using 1--5 exponential terms (as indicated in the legend) applied to the continuous and intermittent ACF decay. These ACFs underpin the HB lifetime analysis discussed in the main text.}
\label{fig_acf_FAPI-faPb-y-FAPI-MAPI}
\end{figure*}

\begin{figure*}[!tbp]
  \centering 
  \begin{minipage}[b]{0.49\textwidth}
    \includegraphics[width=\textwidth]{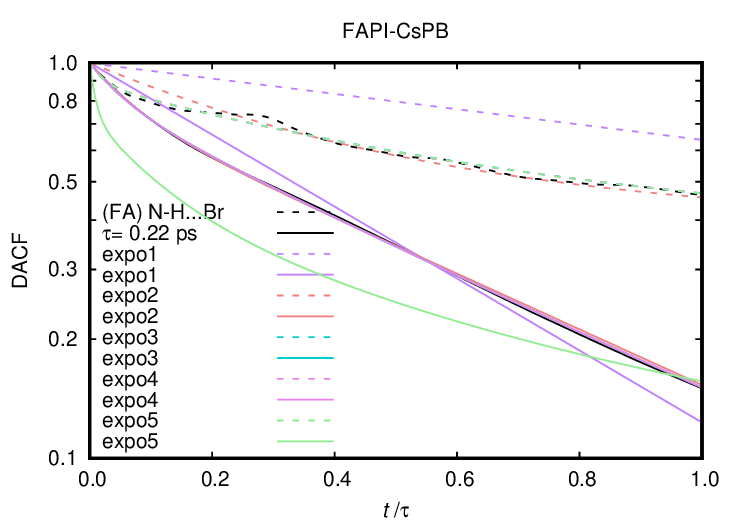}
  \end{minipage}
  \begin{minipage}[b]{0.49\textwidth}
    \includegraphics[width=\textwidth]{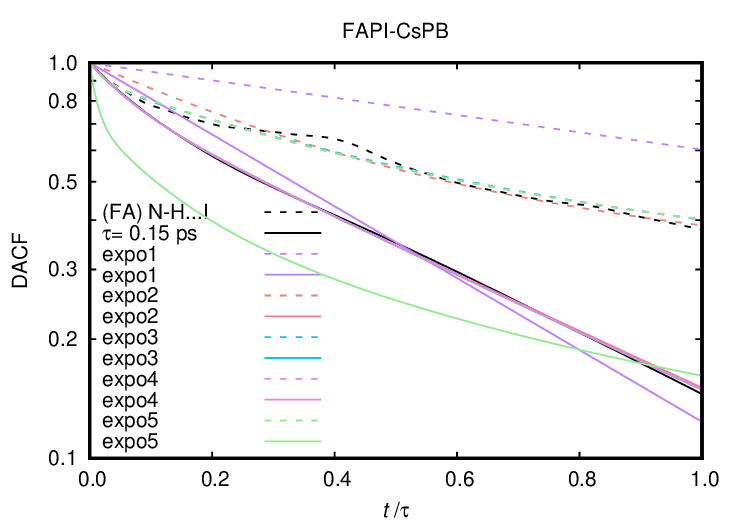}
  \end{minipage}
\caption{Hydrogen-bond autocorrelation functions (ACFs) for \hb{N}{I} and \hb{N}{Br} HBs in \fapicspb{}. Solid line (black): continuous ACF; dotted line (black): intermittent ACF. Poly-exponential fits using 1--5 exponential terms (as indicated in the legend) applied to the continuous and intermittent ACF decay. These ACFs underpin the HB lifetime analysis discussed in the main text.}
\label{fig_acf_FAPI-CsPB}
\end{figure*}

\begin{figure*}[!tbp]
  \centering
  \begin{minipage}[b]{0.40\textwidth}
    \includegraphics[width=\textwidth]{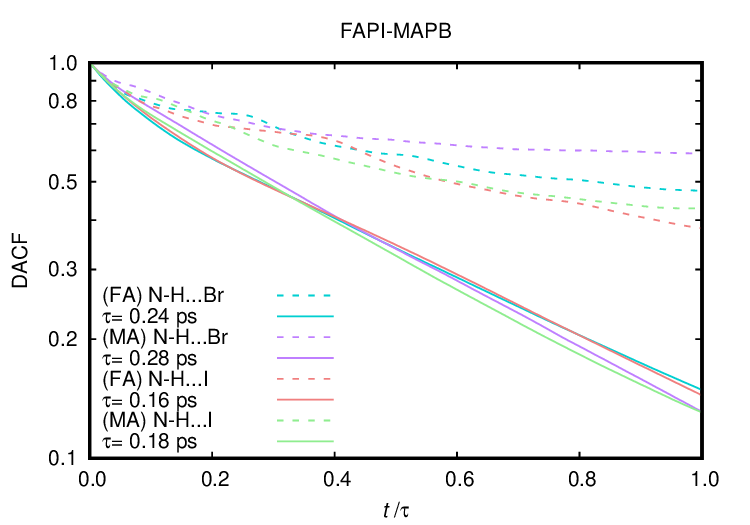}
  \end{minipage}
  \begin{minipage}[b]{0.40\textwidth}
    \includegraphics[width=\textwidth]{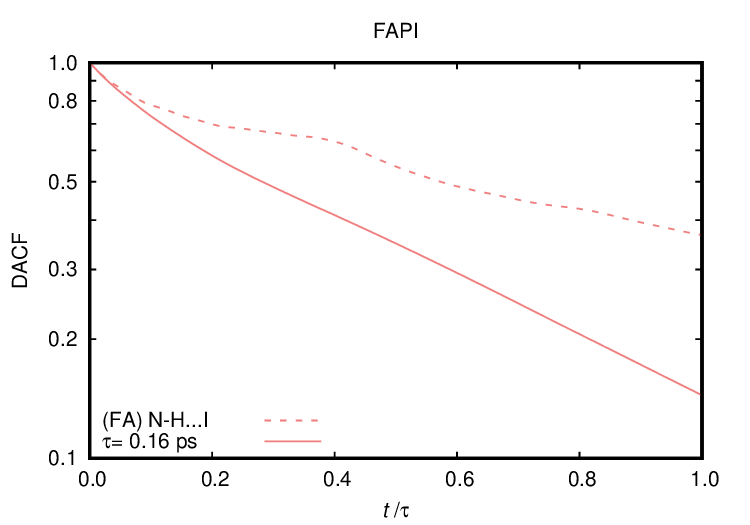}
  \end{minipage}
  \begin{minipage}[b]{0.40\textwidth}
    \includegraphics[width=\textwidth]{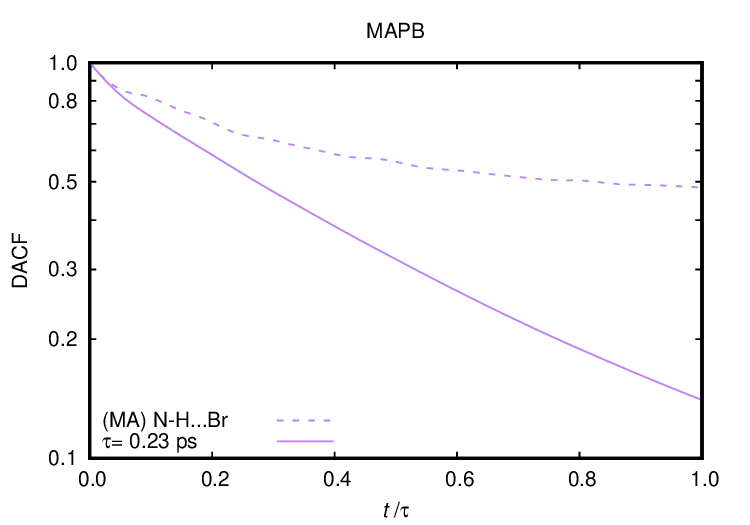}
  \end{minipage}
  \begin{minipage}[b]{0.40\textwidth}
    \includegraphics[width=\textwidth]{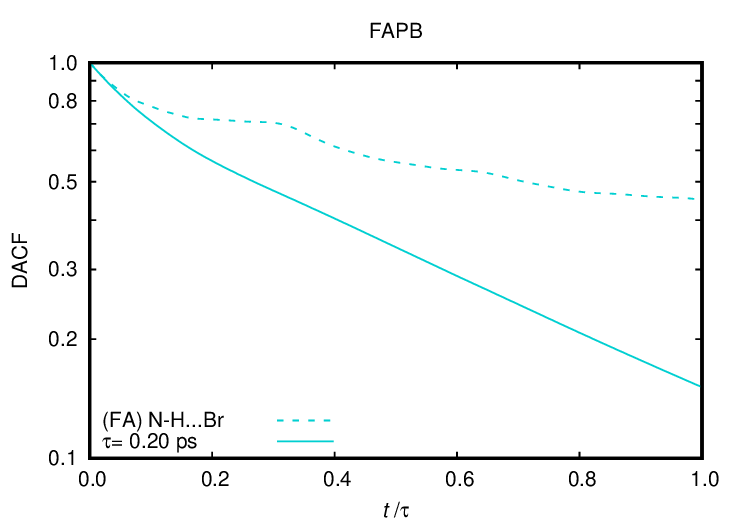}
  \end{minipage}
  \begin{minipage}[b]{0.40\textwidth}
    \includegraphics[width=\textwidth]{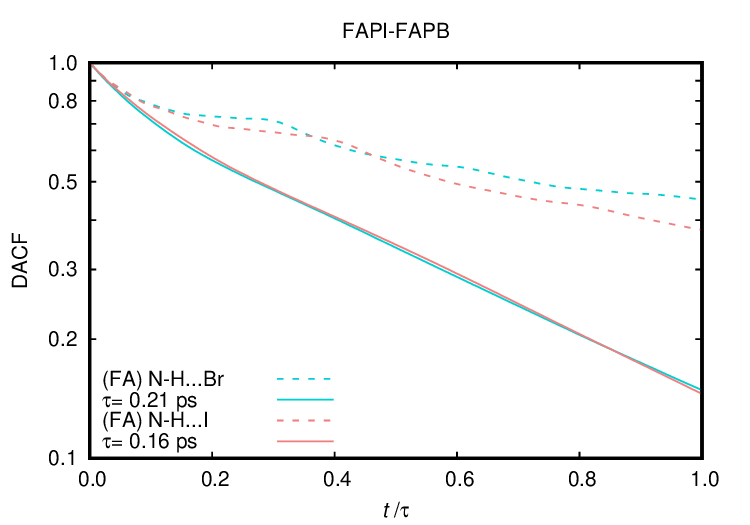}
  \end{minipage}
  \begin{minipage}[b]{0.40\textwidth}
    \includegraphics[width=\textwidth]{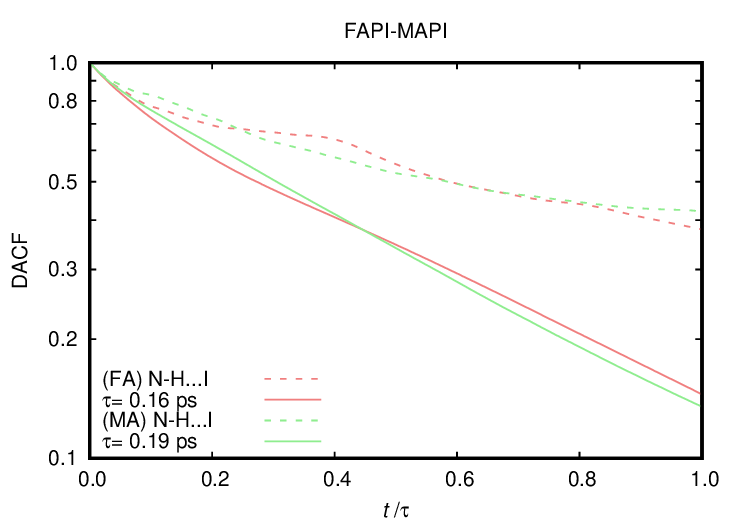}
  \end{minipage}
  \begin{minipage}[b]{0.40\textwidth}
    \includegraphics[width=\textwidth]{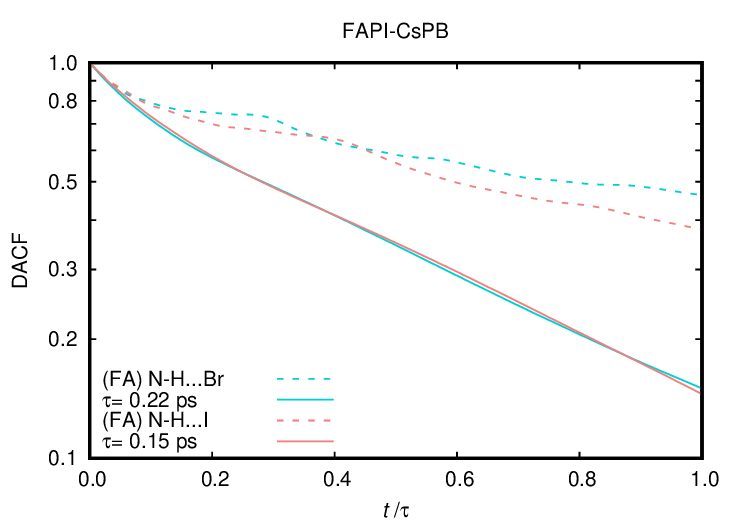}
  \end{minipage}
  \begin{minipage}[b]{0.40\textwidth}
    \includegraphics[width=\textwidth]{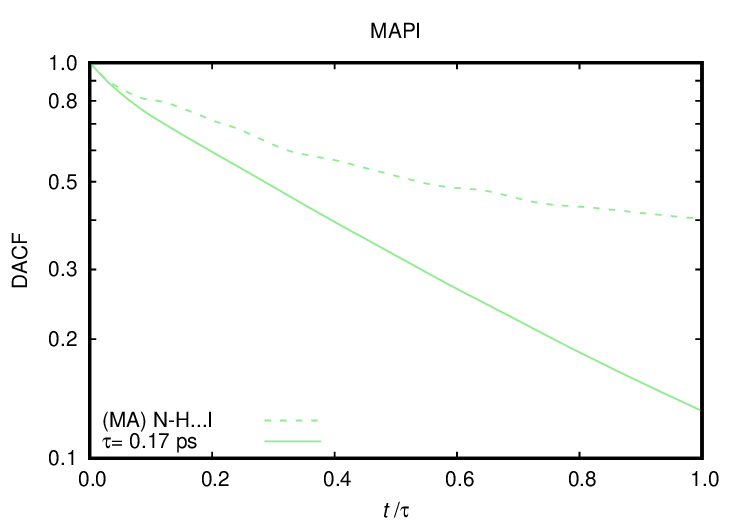}
  \end{minipage}
\caption{Hydrogen-bond autocorrelation functions (ACFs) for all studied systems. Aggregate \hb{N}{I} and \hb{N}{Br} ACFs are shown for \fapimapb{}, \fapi{}, \mapb{}, \fapb{}, \fapifapb{}, \fapimapi{}, \fapicspb{}, and \mapi{}. Solid lines correspond to the continuous ACF (uninterrupted HBs), and dotted lines to the intermittent ACF (allowing HB breaking and reforming).}
\label{fig_acf_all}
\end{figure*}

To complement the main-text discussion of HB lifetimes, Figures~\ref{fig_acf_FAPI-MAPB}--\ref{fig_acf_FAPI-CsPB} report the \hb{N}{I} and \hb{N}{Br} HB autocorrelation functions (ACFs) computed from the AIMD trajectories for representative mixed and pure systems. For each HB type we show both the continuous ACF (solid line), which counts only uninterrupted HBs, and the intermittent ACF (dotted line), which allows HB breaking and reforming. The ACF decays were fitted using sums of exponentials (1--5 terms) to extract characteristic HB lifetimes and to assess the robustness of the fitted time scales with respect to the fitting form. For convenience, Fig.~\ref{fig_acf_all} compiles the resulting ACFs for all systems together.


\section{Root Mean Square Displacement (RMSD) Analysis}

\begin{figure}
    \centering
    \includegraphics[width=0.65\linewidth]{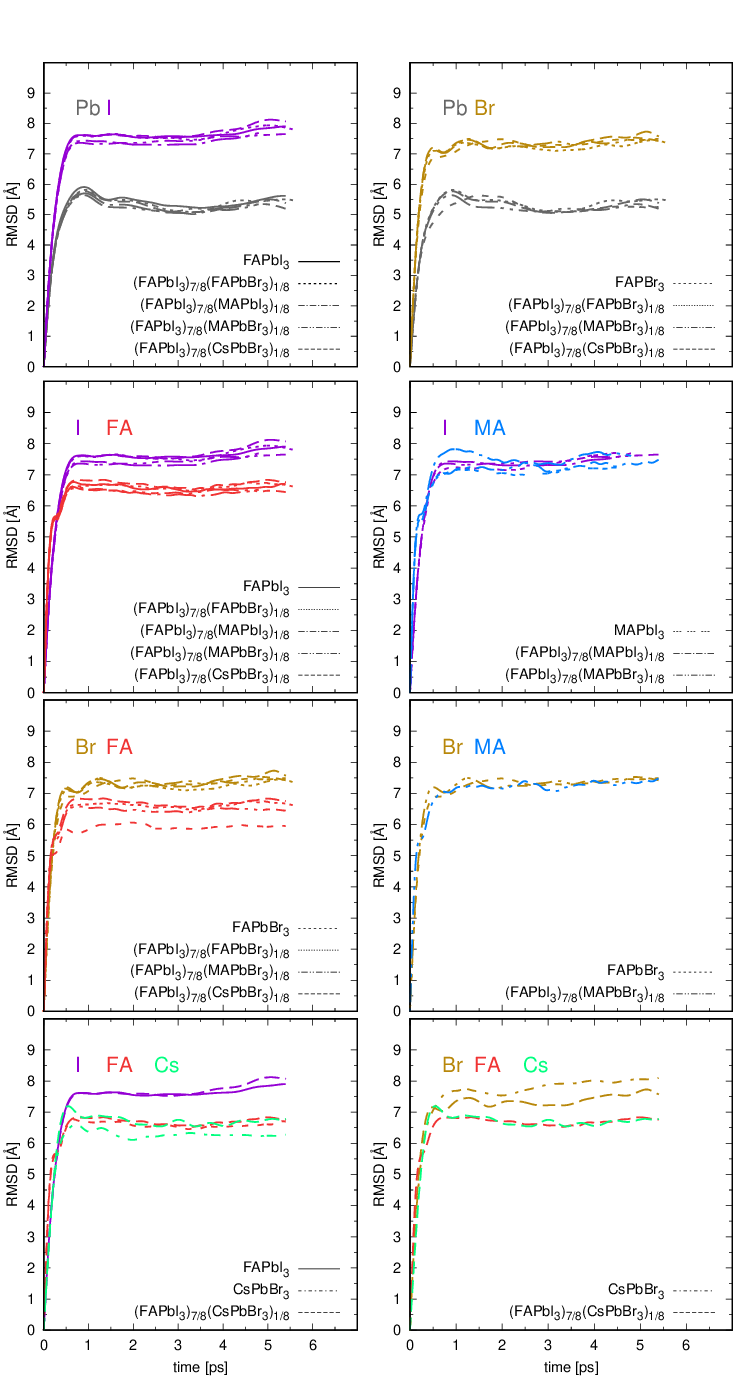}
    \caption{RMSD--time curves grouped by species (Pb, I, Br) and A-site components (FA, MA, Cs) for the pure and mixed perovskite compositions. For FA/MA, RMSD corresponds to the molecular center of mass (COM). All
    curves were computed over the same production trajectory window.}
    \label{fig:rmsd}
\end{figure}

Figure~\ref{fig:rmsd} shows RMSD-time curves reported for atomic species (Pb, I, Br) and by A-site components (FA, MA, Cs), for the pure and mixed perovskite compositions. For molecular cations, RMSD was computed for the
center of mass (COM), whereas for inorganic species it was computed for the atomic positions. These comparisons were used to examine whether local mobility trends correlate with the thermodynamic signatures discussed in the main text.

Across all compositions, the dynamics of the Pb sublattice are essentially unchanged by mixing, indicating that mixing does not disrupt the inorganic cage on the timescale probed by AIMD. Iodide motion is also nearly invariant across the studied alloys. In MA-containing systems, the RMSD of the MA COM closely follows that of iodide, consistent with stronger and more persistent MA--I interactions. In contrast, the FA COM RMSD deviates more substantially from that of I, consistent with weaker and less persistent FA--I hydrogen bonds and greater orientational freedom. This dynamical asymmetry is consistent with the heterogeneous reorientational response observed in FA/MA mixtures and highlights that local mobility differences do not map trivially onto the mixing enthalpy, and may contribute to the rotational-entropy penalty discussed in the main text.

In the Br--FA panel, the FA COM RMSD is smaller in pure \fapb{} than in the Br-containing mixed compositions, whereas the RMSD of Br remains nearly invariant across the same set of systems. This suggests that FA is more
spatially confined in a homogeneous Br environment, while I/Br chemical disorder leads to a more heterogeneous local environment and thus to larger FA displacements. Importantly, this enhanced FA mobility in the mixed systems does not correlate directly with their thermodynamic stability, supporting the conclusion that hydrogen-bond dynamics are not the primary driver of alloy stabilization.

In \fapicspb{}, the FA COM and Cs RMSD curves have comparable amplitudes, indicating that the two A-site species coexist dynamically within the same inorganic framework. This contrasts with FA--MA mixtures, where differences in size, geometry, and interaction strengths can lead to a more heterogeneous A-site dynamical response. Importantly, the comparable mobility of FA and Cs does not imply a reduced enthalpy of mixing; rather, \fapicspb{} remains thermodynamically favorable at 350~K because configurational entropy dominates the free-energy balance despite a positive $\Delta H_\text{mix}$, consistent with the thermodynamic analysis in the main text.

\textcolor{black}{
\section*{References}
\begin{enumerate}[label=(\arabic*)]
    \item Pitzer, K. S.; Gwinn, W. D. Energy Levels and Thermodynamic Functions for Molecules with Internal Rotation I. Rigid Frame with Attached Tops. \emph{J. Chem. Phys.} \textbf{1942}, 10 (7), 428 -- 440.
    \item Brehm, M.; Thomas, M.; Gehrke, S.; Kirchner, B. TRAVIS—A Free Analyzer for Trajectories from Molecular Simulation. \emph{J. Chem. Phys.} \textbf{2020}, 152 (16), 164105.
\end{enumerate}
}